\documentclass[opre,nonblindrev,copyedit]{informs2}

\usepackage{endnotes}
\let\footnote=\endnote

\usepackage{natbib}
 \bibpunct[, ]{(}{)}{,}{a}{}{,}%
 \def\bibfont{\small}%
 \def\newblock{\ }%

 \usepackage{color}
\usepackage{psfrag}
\usepackage{dsfont}

\TheoremsNumberedThrough     
\ECRepeatTheorems

\EquationsNumberedThrough    

\begin{document}



\RUNTITLE{Park et al.}

\TITLE{Bounding Procedures for Stochastic Dynamic Programs with Application to the Perimeter Patrol Problem}

\ARTICLEAUTHORS{%
\AUTHOR{Myoungkuk Park}
\AFF{Department of Mechanical Engineering, Texas A\&M University, College Station, TX 77843, \EMAIL{robotian@gmail.com}}

\AUTHOR{Krishnamoorthy Kalyanam}
\AFF{Infoscitex Corporation, Dayton, OH 45431, \EMAIL{krishna.kalyanam@gmail.com}}

\AUTHOR{Swaroop Darbha }
\AFF{Department of Mechanical Engineering, Texas A\&M University, College Station, TX 77843, \EMAIL{dswaroop@tamu.edu}}

\AUTHOR{Phil Chandler}
\AFF{Control Design \& Analysis Branch, Air Force Research Laboratory, WPAFB, OH 45433, \EMAIL{phillip.chandler@wpafb.af.mil}}
 
\AUTHOR{Meir Pachter}
\AFF{Electrical Engineering Department, Air Force Institute of Technology, WPAFB, OH 45433, \EMAIL{meir.pachter@afit.edu}}
} 

\ABSTRACT{
One often encounters the curse of dimensionality in the application of dynamic programming to determine optimal policies for controlled Markov chains. In this paper, we provide a method to construct sub-optimal policies along with a bound for the deviation of such a policy from the optimum via a linear programming approach. The state-space is partitioned and the optimal cost-to-go or value function is approximated by a constant over each partition. 
By minimizing a non-negative cost function defined on the partitions, one can construct an approximate value function which also happens to be an upper bound for the optimal value function of the original Markov Decision Process (MDP). As a key result, we show that this approximate value function is {\it independent} of the non-negative cost function (or state dependent weights as it is referred to in the literature) and moreover, this is the least upper bound that one can obtain once the partitions are specified. Furthermore, we show that the restricted system of linear inequalities also embeds a family of MDPs of lower dimension, one of which can be used to construct a lower bound on the optimal value function. The construction of the lower bound requires the solution to a combinatorial problem. We apply the linear programming approach to a perimeter surveillance stochastic optimal control problem and obtain numerical results that corroborate the efficacy of the proposed methodology.
}%



\maketitle

%
\section{Introduction}
\label{sec:intro}

The Linear Programming (LP) approach to solving dynamic programs (DPs) originated from the papers: \cite{Manne60,Epenoux63, Denardo70, Hordijk79}. 
The basic feature of an LP approach for solving DPs corresponding to maximization of a discounted payoff 
is that the optimal solution of the DP (also referred to as the optimal value function) is the optimal solution of the LP for {\it every} non-negative cost function. The constraint set describing the feasible solution of the LP and the number of independent variables are typically very large ({\it curse of dimensionality}) and hence, obtaining the exact solution of a DP (stochastic or otherwise) via an LP approach is not practical. Despite this limitation, an LP approach provides a tractable method for approximate dynamic programming  \citep{mendelssohn80, Schweitzer85, Trick97} and the advantages of this approach may be summarized as follows:
\begin{enumerate}
 \item One can restrict the value function to be of a certain parameterized form, thereby reducing the dimension of the LP to the size of the parameter set to make it tractable. 
\item The solution to the LP provides upper bounds for the value function (lower bounds, if minimizing a discounted cost, as opposed to maximizing discounted payoff, is considered as the optimization criteria). 
\end{enumerate}
The main questions regarding the tractability and quality of approximate DP revolve around restricting the value function in a suitable way. The questions are: (1) How does one restrict the value function, i.e., what basis functions should one choose for parameterizing the value function? 
(2) Are there any (a posteriori) bounds that one can provide about the value function from the solution of a restricted LP? If the restrictions imposed on the value function are consistent with the physics/structure of the problem, one can expect reasonably tight bounds. There is another question that naturally arises: In the unrestricted case, the optimal solution of the LP is independent of the choice of the non-negative cost function. While it is unreasonable to expect that the optimal value function be a feasible solution of the restricted LP, one can ask if the optimal solution of the restricted LP is the same for {\it every} choice of non-negative cost function for the LP. It has been reported in the literature that this is unfortunately not the case \citep{deFarias03}.

If the LP is not properly restricted, it can lead to poor approximation and perhaps, even infeasibility \citep{Gordon99}. A common approach is to approximate the value (cost-to-go) function by a linear functional of a priori chosen basis functions \citep{Schweitzer85}. This approach is attractive in that for a certain class of basis functions, feasibility of the approximate (or restricted) LP is guaranteed \citep{deFarias03}. 
 A straightforward method for selecting the basis functions is through a state aggregation method. Here the state space is partitioned into disjoint sets or partitions and the approximate value function is restricted to be the same for all the states in a partition. The number of variables for the LP therefore reduces to the number of partitions. State aggregation based approximation techniques were originally proposed by \cite{Axs,Bean,mendelssohn82}. Since then, substantial work has been reported in the literature on this topic (see \cite{Roy06} and the reference therein). In this article, we adopt the state aggregation method. 
 
 Although imposing restrictions on the value function reduces the size of the restricted LP, the number of constraints does not change. Since the number of constraints is at least of the same order as the number of states of the DP, one is faced with a restricted LP with a large number of constraints. An LP with a large number of constraints may be solved if there is an automatic way to separate a non-optimal solution from an optimal one \citep{Grotschel81}; otherwise, one may have to resort to heuristics or settle for an approximate solution. Separation of a non-optimal solution from an optimal one is easier if one has a compact representation of constraints \citep{Morrison99} or if a subset of the constraints that dominate other constraints can easily be identified from the structure of the problem \citep{Krishna10}. 
Heuristic methods include aggregation of constraints, sub-sampling of constraints \citep{deFarias03}, constraint generation methods\citep{Grotschel91, Schuurmans01} and other approaches \citep{Trick93}. 

If the solution of the restricted LP is the same for {\it every} non-negative cost function of the LP, then it suggests that the constraint set for the restricted LP embeds the constraint set for the exact LP corresponding to a reduced order Markov Decision Process (MDP). 
If one adopts a naive approach and ``aggregates'' every state into a separate partition, we obtain the original exact LP and clearly, for this LP, the solution is independent of the non-negative cost function. It would seem reasonable to expect that this would generalize to partitions of arbitrary size and in fact, we prove this to be the case in this article.
One can construct a sub-optimal policy from the solution to the restricted LP by considering the policy that is greedy with respect to the approximate value function \citep{Porteus75}. By construction, the expected discounted payoff for the sub-optimal policy will be a lower bound to the optimal value function and hence, can be used to quantify the quality of the sub-optimal policy. Also the lower bound will be closer to the optimal value function than the approximate value function by virtue of the monotonicity property of the Bellman operator. But the lower bound computation is not efficient since the procedure involved is tantamount to policy evaluation which involves the solution to a system of linear equations of the same size as the state-space. In this work, we have developed a novel disjunctive LP, whose solution can be used to construct a lower bound to the optimal value function.
The contributions of our work may be summarized as follows:
\begin{itemize}
\item If one were to adopt a state aggregation approach, then the solution to the restricted LP is shown to be {\it independent} of the non-negative cost function. 
Moreover, the optimal solution is dominated by every feasible solution to the restricted LP. 
\item We also show that considering alternate LP formulations via lifting of variables or by considering a bigger feasible set via iterated Bellman inequalities \citep{wang2010} does not improve upon the upper bound provided by the restricted LP.
\item A subset of the constraints of the restricted LP can be used for constructing a lower bound for the optimal value function. However, this involves solving a disjunctive LP, which may not be computationally tractable.
\item We demonstrate the use of aggregation based restricted LPs for a perimeter surveillance stochastic control problem. For the application considered here, we show that both the lower bounding disjunctive LP and the upper bounding restricted LP can be solved efficiently since they both reduce to exact LPs corresponding to some lower dimensional MDPs. 
\end{itemize}

The rest of the paper is organized as follows: we provide a general overview of stochastic dynamic programs in section~\ref{sec:sdp} followed by LP preliminaries in section~\ref{sec:LPprelim}. In section~\ref{sec:BndsPart}, we introduce the aggregation method and discuss the restricted LP approach that can be used to approximate the optimal value function. In the same section, we also present a novel disjunctive LP that can be used to compute a lower bound to the optimal value function. We introduce the perimeter alert patrol problem in section~\ref{sec:patrol} and also elaborate on the efficient LP formulations that arise out of the structure in the problem. We corroborate the structure in the perimeter patrol problem via numerical results in section~\ref{sec:results}. Finally, we support the proposed approximation methodology via simulation results in section~\ref{sec:sim}, followed by summary in section~\ref{sec:con}. Supplementary material and lengthy proofs, that have been left out of the main body of the paper, for clarity, have been included in the Appendix.

\section{Stochastic Dynamic Programming}
\label{sec:sdp}
Consider a discrete-time Markov decision process (MDP) with a finite state space $\mathcal S = \{1,2,\ldots,|\mathcal S|\}$. For each state $x\in\mathcal S$, there is a finite set of available actions
${\mathcal U}_x$. 
From current state $x$, taking action $u\in {\cal U}_x$ under the random influence $Y$ results in a reward $R_u(x)$. The system follows some discrete-time dynamics given by:
\begin{equation}
\label{eq:sysdyn}
x(t+1) = f(x(t),u(t),Y(t)),
\end{equation}
where $t$ indicates time. We assume that the random input $Y$ can only take a finite set of values $Y_l;l=0,\ldots,m$ and there is a probability associated with each choice $p_l$. State transition probabilities $P_u(x, y)$ represent, for each pair $(x, y)$ of states and each action $u\in{\mathcal U}_x$, the probability that the next state will be $ y$ given that the current state is $x$ and the current action taken is $u$ i.e., 
\begin{equation}
\label{eq:tprob}
P_u(x,y) = \left\{\begin{array}{l}
0, \mbox{ if } y\neq f(x,u,Y_l) \mbox{ for any } l \in \{0,\ldots,m\},\\
\sum_{j\in\mathcal C} p_j, \mbox{ where } \mathcal C = \{l| y= f(x,u,Y_l)\}.
\end{array}\right.
\end{equation}
Any \textit{stationary} policy, $\pi$, specifies for each state $x \in {\mathcal S}$, a control action $u = \pi(x)$. 
We abuse notation and also write the transition probability matrix associated with policy $\pi$ to be $P_\pi$, where $P_\pi(x,y) = P_{\pi(x)}(x,y)$. Similarly, we express the column vector of immediate payoffs associated with the policy $\pi$ to be ${R}_\pi$, where ${R}_\pi(x) = {R}_{\pi(x)}(x)$. 
We are interested in solving a stochastic control problem, which amounts to selecting a policy that maximizes the infinite-horizon discounted reward of the form,
\begin{equation}
V_\pi(x_0)=
{\mathbf E}\left[\left.\sum^{\infty}_{t=0}{\lambda^tR_\pi(x(t))}\right | x(0)=x_0  \right], \nonumber
\end{equation}
where $\lambda\in[0,1)$ is a temporal discount factor. We obtain the optimal policy by solving Bellman's equation,
\begin{equation}
\label{eq:Bell}
V^*(x)=\max_{u\in\mathcal U_x}\left\{R_u(x)+\lambda\sum_{l=0}^{m}{p_lV^*(f(x,u,Y_l))}\right\}, \forall x\in\mathcal S,
\end{equation}
where, $V^*(x)$ is the optimal value function (or optimal discounted payoff) starting from state $x$. The optimal policy then is given by,
\begin{equation}
\label{eq:OptPol}
\pi^*(x)=\argmax_{u\in\mathcal U_x}\left\{R_u(x)+\lambda\sum_{l=0}^{m}{p_lV^*(f(x,u,Y_l))}\right\}, \forall x\in\mathcal S.
\end{equation}
The Bellman equation (\ref{eq:Bell}) can be solved using standard DP methods such as value iteration \citep{Howard1960dynamic} or policy iteration \citep{bellman1957};
however, it is computationally not tractable, if the size of state space considered is unmanageably large.
For this reason, one is interested in tractable approximate methods that yield suboptimal solutions with some guarantees on the deviation of the associated approximate value function from the optimal one.

\subsection{Linear Programming Approach}
\label{sec:LPprelim} 

In this subsection, we briefly touch upon two lemmas that we will use in the subsequent sections. 
Bellman's equation suggests that the optimal value function satisfies the following set of linear inequalities, which we will refer to as the Bellman inequalities:
\begin{eqnarray}
\label{eq:bellineq}
V(x) &\geq& R_u(x) + \lambda \sum_{l=0}^m p_l V(f(x,u, Y_l)), \; \forall u\in\mathcal U_x,\; \forall x\in\mathcal S.\nonumber\\
\Leftrightarrow V &\geq& R_{u} + \lambda P_{u} V, \; \forall \; u.
\end{eqnarray}
Consider any integer $L \geq 1$ and for $j=1, 2, \ldots, L$, let $V_j$ be a vector satisfying a generalization of the Bellman inequalities, referred to as the iterated Bellman inequalities \citep{wang2010}:
\begin{eqnarray}
\label{eq:GenIBE1u}
V_{j+1}(x) &\geq& R_u(x) + \lambda \sum_{l=0}^m p_l V_j(f(x, u, Y_l)), \; \; \forall \; x, u, \quad \forall j=1, 2, \ldots, L-1, \\
\label{eq:GenIBE2u}
V_1(x) &\geq& R_u(x) + \lambda \sum_{l=0}^m p_l V_L(f(x,u,Y_l)), \; \; \forall \; x, u.
\end{eqnarray}
Clearly, when $L=1$, the above system of inequalities collapses to the Bellman inequalities. 
The iterated Bellman inequalities may be compactly represented as:
\begin{eqnarray}
\label{eq:GenIBEcompact}
V_{j+1} &\geq& R_{u} + \lambda P_{u} V_j, \; \; \forall \; u, \; \; j=1, 2, \ldots, L-1, \nonumber\\
V_1 &\geq& R_{u} + \lambda P_{u} V_L, \; \; \forall \; u.
\end{eqnarray}
We note that the above set of inequalities have cyclic symmetry, i.e., one gets the same set of inequalities by replacing the vectors $V_1, V_2, \ldots, V_L$ by $V_2, V_3, \ldots, V_L, V_1$ respectively. 
Let $\pi$ be any stationary policy. 
Then we have,
\begin{eqnarray}
\label{eq:GenIBE1}
V_{j+1} &\geq& R_{\pi} + \lambda P_{\pi} V_j, \; \; j=1, 2, \ldots, L-1, \\
\label{eq:GenIBE2}
V_1 &\geq& R_{\pi} + \lambda P_{\pi} V_L. 
\end{eqnarray}
By recursively applying (\ref{eq:GenIBE1}) to $V_L,V_{L-1},\ldots$ etc., in (\ref{eq:GenIBE2}), we get,
\begin{eqnarray*}
[I-\lambda^L P_{\pi}^L]V_1 \geq [I+ \lambda P_{\pi} + \cdots + \lambda^{L-1}P_{\pi}^{L-1}]R_{\pi}, \; \;  \forall \; \pi.
\end{eqnarray*}
By cyclic symmetry, every $V_j, \; j= 2, 3, \ldots, L,$ also satisfies the above inequality. 
\begin{lemma}
\label{lem:autoUB1}
Let the vector $V$ satisfy the following set of inequalities:
\begin{eqnarray}
\label{eq:GenIBE}
\left[I-\lambda^LP_{\pi}^L\right]V \geq \left[I+ \lambda P_{\pi} + \cdots + \lambda^{L-1}P_{\pi}^{L-1}\right]R_{\pi}, \quad \forall \; \pi.
\end{eqnarray}
Then, we have $V \geq V^*$.
\end{lemma}
\begin{remark}
\label{rem:UBval}
We readily see that every feasible solution of the system of inequalities (\ref{eq:bellineq})  or (\ref{eq:GenIBEcompact}) is lower bounded by the optimal value function $V^*$. By cyclic symmetry, we conclude that every feasible $V_j, \; j=1, \ldots, L$ is also lower bounded by $V^*$.
\end{remark}
The following result relates the optimal value function to the optimal solution of an LP with a non-negative cost function and constraints of the form given by the Bellman inequalities (\ref{eq:bellineq}) or iterated Bellman inequalities (\ref{eq:GenIBEcompact}). 
\begin{lemma} 
\label{lem:LPopt1}
Let ${c}$ be a vector of state-dependent weights with $c(x) \geq 0$ for every $x\in\mathcal S$. Then $V^*$ minimizes the linear functional $ c^TV$ among all $V$'s satisfying the Bellman inequalities (\ref{eq:bellineq}). Correspondingly, the $L$-tuple $(V^*, \cdots, V^*)$ minimizes the linear functional 
$\sum_{j=1}^L c^TV_j$ among all $L$-tuples $(V_1, \ldots, V_L)$ satisfying the iterated Bellman inequalities (\ref{eq:GenIBEcompact}).
\end{lemma}
\proof{Proof of Lemma~\ref{lem:LPopt1}.} 
The proof follows from the fact that $V \geq V^*$ and hence, $c^T(V-V^*) \geq 0$. Since $V^*$ is feasible for the inequalities (\ref{eq:GenIBEcompact}) for any $L\geq 1$, the result follows. Similarly, since the $L$-tuple
$(V^*, \cdots, V^*)$ is feasible for (\ref{eq:GenIBEcompact}) and since $V_j \geq V^*$ for $j=1, 2, \ldots, L$, it readily follows that the $L$-tuple is optimal.
\Halmos 
\endproof
\section{Bounds using Partitioning}
\label{sec:BndsPart}

Let the set of all states $\mathcal{S}$ be partitioned into $M$ disjoint sets, $\mathcal{S}_i,i=1,\ldots,M$. We will call the set $\mathcal{S}_i$ the $i^{\text{th}}$ partition. 
Henceforth, we will use the following notation: if $f(x,u,Y_s)$ represents the state the system transitions to starting from $x$ and subject to a control input $u$ and a stochastic disturbance $Y_s$, then $\bar{f}(x,u,Y_s)$ represents the partition to which the final state belongs. 
For a given $u$ and partition index $i$, we define the tuple $z^{i,u}_x = (\bar{f}(x,u,Y_0),\bar{f}(x,u,Y_1),\hdots,\bar{f}(x,u,Y_m))$ for every $x\in\mathcal S_i$. We denote by $\mathcal{T}(i,u)$ the set of all distinct $z^{i,u}_x$ for a given  partition index $i$ and control $u$. 

\subsection{Restricted Linear Program}
\label{sec:resLP}
We have, from Lemma~\ref{lem:LPopt1}, that the optimal solution to the following LP, 
\begin{eqnarray}
\label{eq:exactLP}
ELP&:=&\min {c}^T{V}, \quad \mbox{subject to}\\
 V&\geq& R_u+\lambda P_uV,\quad\forall u,\nonumber
\end{eqnarray}
referred to as the ``exact LP'' in the literature, is the optimal value function $V^*$.
Let us start with restricting the exact LP by requiring further that $V(x)=v(i)$ for all $x\in \mathcal{S}_i$,
$i=1,\ldots,M$. Augmenting these constraints to the exact LP, one gets the following restricted LP.
\begin{eqnarray}
\label{eq:RLP}
RLP&:=&\text{min}\sum_{i=1}^{M}{\sum_{x\in \mathcal{S}_i}{c(x)v(i)}} \quad \mbox{subject to}\quad\\
v(i)&\ge & R_u(x) + \lambda \sum_{l=0}^{m}{p_lv(\bar f(x,u,Y_l))},
\quad\forall x\in \mathcal{S}_i,\: i=1,\ldots,M,\: \forall u.\nonumber
\end{eqnarray}
The restricted LP can also be written in the following compact form:
\begin{eqnarray}
\label{eq:RLPcomp}
RLP&=&\min c^T\Phi v\quad \mbox{subject to}\quad\\
\Phi v&\ge & R_u + \lambda P_u\Phi v,\quad\forall u,\nonumber
\end{eqnarray}
where the columns of $\Phi$ (commonly referred to as ``basis functions'' in the literature) are given by,
\begin{equation}
\label{eq:basis}
\Phi(x,i) = \left\{
\begin{array}{l} 1, \mbox{ if } x\in\mathcal S_i,\\
		        0, \mbox{ otherwise. }\end{array}	
\right.,\quad i=1,\ldots,M.
\end{equation}
The restricted LP typically deals with a much smaller number of variables i.e., $M<<|\mathcal S|$. An approximate value function can be constructed from every feasible solution to $RLP$ according to $V_{up}=\Phi v\Rightarrow V_{up}(x)=v(i)$, $\forall x\in \mathcal{S}_i, i=1,\ldots,M$. 
Since the approximate value function satisfies, by
construction, the Bellman inequalities (\ref{eq:bellineq}),
it is automatically an upper bound to $V^*$ by Lemma~\ref{lem:autoUB1}. So, if ${v^*}$ is the optimal solution to $RLP$ (\ref{eq:RLP}), then clearly, $\Phi v^* \geq V^*$. Now we are ready to address one of the main results of the paper. 
\begin{theorem}
\label{th:RLPindcost}
The optimal solution, $v^*$, to the $RLP$ is independent of the cost vector $c$ once the partitions are specified. 
\end{theorem}
\proof{Proof of Theorem~\ref{th:RLPindcost}.} 
The main idea behind the proof is the following: The constraints in the restricted LP (\ref{eq:RLP}) do not, in general, correspond to those of a Markov Decision Process (MDP) because the transition from one partition to another for a given control $u$ and random input $Y_l$ is not specified unambiguously. This is because different states in the same partition can transition to different partitions for the same $u$ and $Y_l$. If one were to think of a ``random'' selector for a state in a partition, then the specification of $u$, $Y_l$ together with the random selector specifies exactly which partition the system would transition to next, from the current partition. Let us specify the probability of picking a state in a partition, corresponding to the random selector, via the optimal dual variables for $RLP$. 
For a given partition index $i$, the $RLP$ specifies a constraint on $v(i)$ for each $x\in\mathcal S_i$ and $u$. Let the dual variable corresponding to this constraint be $\mu_u^i(x) \geq 0$  and the corresponding optimal dual variable be $\bar{\mu}_u^i(x)$. With this definition, we can proceed to prove the result via the following steps:
\begin{enumerate}
\item We show that for every partition index $i$, there is a $u$ such that $\bar{\mu}_u^{i}(x) > 0$ for some $x \in {\mathcal S}_i$. This is necessary for constructing a MDP of reduced dimension in the next step; otherwise, the corresponding value of $v(i)$ is not lower bounded. 
\item 
We define a reduced order MDP on the partitions with immediate reward and transition probability given by,
  \begin{equation*}
    \label{eq:randsel1}
r_u(i) = \sum_{x \in {\cal S}_i}h_u^i(x)R_u(x)\mbox{ and }  \tilde P_u(i,j) :=\left\{\begin{array}{ll}
\sum_{x \in {\cal S}_i}h_u^i(x) \sum_{y\in\mathcal S_j}P_u(x,y), &\mbox{ if } u\in\mathcal U_i,\\
0, &\mbox{ otherwise, }\end{array}\right.
  \end{equation*}
where $u\in\mathcal U_i$ if  $\sum_{x \in {\cal S}_i} \bar\mu_u^i(x)>0.$ We may interpret the term $h_u^i(x)=\frac{\bar \mu_u^i(x)}{\sum_{x \in {\cal S}_i} \bar\mu_u^i(x)}$ as the probability of 
picking the state $x$ from the partition ${\mathcal S}_i$. 
\item We show that the so-called ``surrogate LP'' obtained by aggregating the constraints of $RLP$ via the optimal dual variables,
\begin{eqnarray}
\label{eq:SLPMDP}
SLP(\bar \mu) &:=& \min \sum_{i=1}^M \underbrace{\sum_{x \in {\mathcal S}_i} c(x)}_{\bar c(i)} v(i), \quad\mbox{subject to}\quad\\
\nonumber
v(i)  &\geq& r_u(i) + \lambda \sum_{j=1}^M \tilde P_u(i,j)v(j), \; \; \forall u\in\mathcal U_i,i=1,\ldots,M,
\end{eqnarray}
is the exact LP corresponding to the reduced order MDP defined in step 2 above.
In essence, for a given $c$, the optimal value function of the reduced order MDP is the optimal solution of  $RLP$. We use the properties of surrogate duality \citep{greenberg1970surrogate, glover1975surrogate, glover1968surrogate} to demonstrate that $SLP(\bar\mu)=RLP$.
\item Finally, to show that the optimal solution to $RLP$ is independent of $c$, we note that the constraints of $SLP(\bar\mu)$ are obtained by taking convex combinations of the constraints in $RLP$. Hence, any feasible solution to $RLP$ is also feasible for $SLP(\bar\mu)$. Since every feasible solution of the exact LP corresponding to an MDP dominates the optimal solution (from Lemma~\ref{lem:autoUB1}), we conclude that the optimal solutions corresponding to two different cost functions $c_1$ and $c_2$ necessarily dominate each other and hence, have to be the same. \Halmos 
\end{enumerate}
\endproof

We shall now establish the surrogate LP result via the following lemma with the proof provided in the Appendix.
\begin{lemma} 
\label{lem:surrLP}
Consider a surrogate LP for the $RLP$ through a set of dual variables, $\mu$ given by:
\begin{eqnarray}
\label{eq:SLP}
SLP(\mu) &:=& \min  \bar c^T v,  \quad \mbox{subject to}\quad \\
\nonumber
\sum_{x \in {\cal S}_i} \mu_u^i(x) v(i) &\geq&  \sum_{x \in {\cal S}_i} 
\mu_u^i(x) \left[R_u(x) + \lambda \sum_{l=0}^{m} p_l v({\bar f(x,u,Y_l)})\right], \;  \forall u,\;  i=1,\ldots,M.
\end{eqnarray}
Then, $\exists\bar \mu \geq 0$ such that,
$SLP(\bar \mu) = RLP,$ and, for every partition index $i=1,\ldots,M$, $\exists u$ such that $\sum_{x \in {\cal S}_i}\bar \mu_u^i(x) > 0$. Moreover, the optimal solution $v^*$ to $RLP$ is independent of the cost vector $\bar c$ and any other feasible solution $v$ to $RLP$ dominates $v^*$.
 \end{lemma}
 
Theorem~\ref{th:RLPindcost} implies that the upper bound for the optimal value function cannot be improved by changing the cost function from a linear to a non-linear function or by restricting the feasible set of $RLP$ further since the optimal solution of $RLP$ is dominated by every feasible solution of $RLP$. Also $\Phi v^*$ is the least upper bound to the optimal value function $V^*$ since any other feasible $v$ to $RLP$ satisfies $\Phi v\geq\Phi v^*$. Hence, a refinement of the upper bound must necessarily involve an enlargement of the feasible set if one wants to stick to an LP formulation, i.e., it should include the feasible set of (\ref{eq:RLP}) and possibly other tighter upper bounds than the optimal solution of $RLP$. Lifting of variables is one way to improve the bound; in this connection, we show in the following section that neither a general lifted LP nor one obtained by including the iterated Bellman inequalities in the constraint set improves the upper bound. 
\begin{remark}
If one considers the sub-optimal dual variables, $\mu_u^i(x) =\frac{1}{|\mathcal S_i|}, \forall x\in\mathcal S_i,\forall u$, then solving the corresponding surrogate dual, $SLP(\mu)$, to obtain an approximate value function, would result in the so-called ``hard aggregation'' method (see Sec. 4 of  \cite{Bertsekas}).
\end{remark}
\begin{remark}
When $\mu$ and $\Phi$ are allowed to have arbitrary positive entries satisfying $\sum_{x=1}^{|\mathcal S|} \mu_u^i(x)= 1, \forall i\in\{1,\ldots,M\}$ and $\sum_{j=1}^M \Phi(y,j) = 1, \forall y \in\mathcal S$, the method is referred to as ``soft aggregation'' \citep{Singh}. Unfortunately, in this case, the optimal solution to the restricted LP formulation (\ref{eq:RLPcomp}) has been shown to be dependent on the cost function \citep{deFarias03}. 
\end{remark}

\subsection{Lifted Restricted Linear Programs}
\label{sec:liftedRLP}
It may appear that we can get tighter upper bounds than those provided by the $RLP$ by considering either lifted LPs whose feasible set is larger than that of $RLP$ or LPs with a different objective function. We will show, in this section, that unfortunately this is not the case. 
In general, one can construct a lifted LP of the form:
\begin{eqnarray}
\label{eq:LLP}
LLP &:=& \min \bar c^T v + d^Tz, \nonumber\quad \mbox{subject to}\quad\\
\label{eq:LLP1}
V(x)&\ge& R_u(x) + \lambda\sum_{l=0}^{m}{p_l(V(f(x,u,Y_l)))},\: \forall x,u,  \\
\label{eq:LLP2}
V(x)&=&v(i),\: \forall x\in \mathcal{S}_i, \: i=1,\ldots,M, \\
\nonumber
z&\geq&0,
\end{eqnarray}
where $z$ is the additional vector of variables used in lifting so that the feasible set is not empty. 
Then, it follows that if  $(\tilde{v}, \tilde z)$ is optimal to $LLP$, then $\tilde{v}$ will be a feasible solution to $RLP$. Consequently,  $\Phi\tilde{v}\geq\Phi v^*$, where $v^*$ is the optimal solution of the $RLP$. In other words, one gets no better bound via lifting if the constraints (\ref{eq:LLP1}) and (\ref{eq:LLP2}) are included. One could also use the iterated Bellman inequalities (\ref{eq:GenIBEcompact}) for constructing a lifted LP of the form: 
\begin{eqnarray}
\label{eq:IBEObj}
IB &:=& \min \sum_{j=1}^L\bar  c^T v_{j}, \quad \mbox{subject to}\quad \nonumber\\
\label{eq:IBEcon1}
v_{j+1}(i) &\geq& R_u(x) + \lambda \sum_{l=0}^m p_l v_{j}({\bar f(x,u,Y_l)}), \quad \forall x \in \mathcal{S}_i, \; \; \forall i, u, \; \; j = 1, \ldots, L-1, \\
\label{eq:IBEcon2}
v_1(i) &\geq& R_u(x) + \lambda \sum_{l=0}^m p_l v_L({\bar f(x,u,Y_l)}), \quad \forall x \in \mathcal{S}_i, \; \; \forall i, u.
\end{eqnarray}
Again, it turns out that the above lifted $LP$ is incapable of providing a better bound, as can be seen from the following result.
\begin{theorem}
\label{th:IBEindc}
If $v_{IB} = (v_1, \cdots, v_L)$ is a feasible solution to $IB$, then $v_{j} \geq v^*$ for $j =1, \ldots, L$, where $v^*$ is the optimal solution to $RLP$.
\end{theorem}
The proof for Theorem~\ref{th:IBEindc} follows along the lines of Lemma~\ref{lem:surrLP}. We will construct a surrogate LP for the lifted LP (\ref{eq:IBEObj}) with the optimal dual variables of $RLP$. We immediately recognize that the inequalities defining the surrogate LP are, in fact, the iterated Bellman inequalities associated with the reduced order MDP defined in step 2 of the proof of Theorem~\ref{th:RLPindcost}. So, the result follows from Lemma~\ref{lem:LPopt1} and Remark~\ref{rem:UBval}.
\proof{Proof of Theorem~\ref{th:IBEindc}.} 
Let $\bar \mu$ be the optimal dual variables to $RLP$ (\ref{eq:RLP}). From Lemma~\ref{lem:surrLP}, for every partition index $i\in\{1,\ldots,M\}$, there exists a $u$ such that $\sum_{x \in \mathcal{S}_i} \bar \mu_u^i(x)>0$.
For a fixed $i$ and $u$, we multiply the inequalities (\ref{eq:IBEcon1}, \ref{eq:IBEcon2}) associated with a particular $x \in \mathcal{S}_i$ with $\bar\mu_u^i(x)$ and sum over all the $x \in \mathcal{S}_i$.
Then, we get the following surrogate LP:
\begin{eqnarray}
\label{eq:SurrIBE}
SIB &:=&  \; \; \min  \sum_{j=1}^L \bar  c^T v_{j}, \quad \mbox{subject to}\quad \nonumber\\
\label{eq:SurrIBEcon1}
v_{j+1}(i) &\geq& r_u(i) + \lambda \sum_{x \in \mathcal{S}_i} h_u^i(x)\sum_{l=0}^m p_lv_{j}({\bar f(x,u,Y_l)}), \; \; \forall u\in\mathcal U_i, \forall i, \; \; j = 1, \ldots, L-1, \\
\label{eq:SurrIBEcon2}
v_1(i) &\geq& r_u(i) + \lambda \sum_{x \in \mathcal{S}_i} h_u^i(x)\sum_{l=0}^m p_lv_L({\bar f(x,u,Y_l)}), \; \; \forall u\in\mathcal U_i, \forall i,
\end{eqnarray}
where, $u\in\mathcal U_i$ if $\sum_{x \in {\cal S}_i} \bar \mu_u^i(x) > 0$.
As before, the one-step reward function,
\[r_u(i) = \frac{\sum_{x \in {\cal S}_i} \bar \mu_u^i(x)R_u(x)}{\sum_{x \in {\cal S}_i} \bar\mu_u^i(x)},
\mbox{ where, }
h_u^i(x) = \frac{ \bar \mu_u^i(x)}{\sum_{x \in \mathcal{S}_i} \bar \mu_u^i(x)}, \quad \forall u\in\mathcal U_i.
\]
By Lemma~\ref{lem:LPopt1}, the optimal solution to $SIB$ is of the form ${v}_{SIB}^* = ({v}^*, \cdots, {v}^*)$, where
$v^*$ is the optimal solution to $SLP(\bar \mu)$ (and by Lemma~\ref{lem:surrLP}, also the optimal solution to $RLP$). 
Since any feasible solution  to $IB$, $v_{IB}=\{v_1,\ldots,v_L\}$ is also feasible to $SIB$, it follows, from Lemma~\ref{lem:autoUB1}, that $v_{j}\geq v^*$ for every $j=1,\dots,L$.
\Halmos 
\endproof
So, we conclude that lifting through the use of iterated Bellman inequalities does not help in finding a tighter upper bound than the $RLP$ optimal solution. Also using any other non-linear objective function will not improve the upper bound as long as the iterated Bellman inequalities (\ref{eq:IBEcon1}) and (\ref{eq:IBEcon2}) are included in the constraints set. In the next section, we focus our attention on the construction of a lower bound for the optimal value function.

\subsection{Lower Bound for the Optimal Value Function}
\label{sec:LowerBnd}
For any candidate approximate value function $\tilde V$, one can construct a sub-optimal ``greedy'' policy according to: 
$$\tilde\pi(x) = \argmax_u \left\{R_u(x) + \lambda \sum_y P_u(x,y)\tilde V(y)\right\}, \quad\forall x\in\mathcal S. $$
Let us define the improvement in value function, $\tilde\alpha(x) := R_{\tilde\pi}(x) + \lambda \sum_y P_{\tilde\pi}(x,y)\tilde V(y) - \tilde V(x)$. Note that there is no improvement, i.e., $\tilde\alpha\equiv0$, when $\tilde V=V^*$. The expected discounted payoff, $V_{\tilde\pi}$, corresponding to the suboptimal policy $\tilde \pi$,
satisfies the following bound \citep{Porteus75}:
$$\tilde V(x) + \frac{1}{1-\lambda} \min_{y}\tilde\alpha(y)\leq V_{\tilde\pi}(x) \leq V^*(x), \quad\forall x\in\mathcal S. $$
In our experience, the lower bound to the optimal value function provided by $V_{\tilde\pi}$ is very conservative. Also computation of $V_{\tilde\pi}$ involves solving a linear system of equations of size $|\mathcal S|$, which would be expensive for a large state-space. So, we construct a novel alternate lower bound as follows.
Recall that for each $x \in {\cal S}_i$, $V^*(x)$ satisfies the Bellman inequality (\ref{eq:bellineq}):
\begin{eqnarray}
V^*(x) &\geq& R_u(x) + \lambda \sum_{l=0}^{m} p_l V^*(f(x,u,Y_l)), \quad\forall u,\nonumber\\
\label{eq:minValfn}
&\geq& R_u(x) + \lambda \sum_{l=0}^{m}p_l \min_{y\in\bar f(x,u,Y_l)}V^*(y), \quad\forall u.
\end{eqnarray}
Let $\bar w(i) := \min_{x \in {\cal S}_i} V^*(x),\;\;i=1,\ldots,M$.
Then, it follows from (\ref{eq:minValfn}) that,
\begin{eqnarray}
\label{eq:NLPmot}
\bar w(i) &\geq& \min_{x \in {\cal S}_i} \left\{R_u(x) + \lambda \sum_{l=0}^{m} p_l \bar w({\bar f(x,u,Y_l)})\right\}\quad\forall u,\;i=1,\ldots,M.
\end{eqnarray}
The above set of inequalites motivates the following non-linear program:
\begin{eqnarray}
NLP &:=& \min \bar c^T w, \quad\mbox{subject to}\quad\nonumber\\
\label{eq:NLPcon}
 w(i)&\geq& \min_{x \in {\cal S}_i} \left\{R_u(x) + \lambda \sum_{l=0}^{m} p_l w({\bar f(x,u,Y_l)})\right\}, \quad \forall u,\; i=1,\ldots,M.
 \end{eqnarray}
Let $w^*$ be the optimal solution to $NLP$.  By construction, we see that $\bar w$ is a feasible solution to the $NLP$ and hence,
$$\bar c^Tw^* \leq \bar c^T\bar w =\sum_{i=1}^M \bar c(i) \min_{x\in {\cal S}_i} V^*(x).$$
So, by choosing $\bar c(i) = 1$ and $\bar c(j) = 0$ for all $j \neq i$, one can obtain a lower bound to the optimal value function for all the states in the $i^{th}$ partition. Moreover, if the problem under consideration exhibits a special structure, one can show that $NLP$ collapses to an LP that can be efficiently solved. The perimeter patrol problem considered herein exhibits such a structure; we demonstrate this in the next section.
\begin{remark}
The $NLP$ is referred to as a disjunctive linear program \citep{balas1979} and the optimal solution to $NLP$ is the solution that minimizes the same linear objective function over the convex hull  of the feasible solutions of $NLP$. \cite{balas1998disjunctive} provides two methods to solve the problem: one through a lifted representation for the convex hull of the feasible set of $NLP$ and the other through a cutting plane technique. 
Since the number of lifted variables is of $O(M^2|{\cal U}|)$; if $M= 10,000$, then one must deal with a lifted LP with 100 million variables. The original (non-aggregated LP) has about $10$ million variables and hence, the lifted representation method is not practical. For this reason, the cutting plane technique is a viable alternate method.
\end{remark}
\begin{remark}
The lower bound provided by $NLP$ is a non-trivial one because the optimal solution is the optimal value function of a reduced order MDP. Hence, the lower bound will be better than at least the value function associated with some suboptimal policy and so, is non-trivial and non-conservative. 
\end{remark}
\begin{remark}  While ${\cal S}_i$ may have a lot of states, the number of entries on the right hand side of the non-linear constraint (\ref{eq:NLPcon}) over which the minimization must be carried out is the cardinality of ${\cal T}(i,u)$. $NLP$ is combinatorial in nature, in the sense that one must pick one $(m+1)$ tuple for each $i$ and $u$ over which the optimization must be carried out. However, for each $(m+1)$ tuple picked, one obtains an MDP. So, the system of inequalities (\ref{eq:NLPcon}) describes a family of underlying MDPs.
\end{remark}

\section{Perimeter Patrol Problem}
\label{sec:patrol}
\begin{figure}
\FIGURE
	{\includegraphics*[width=3in]{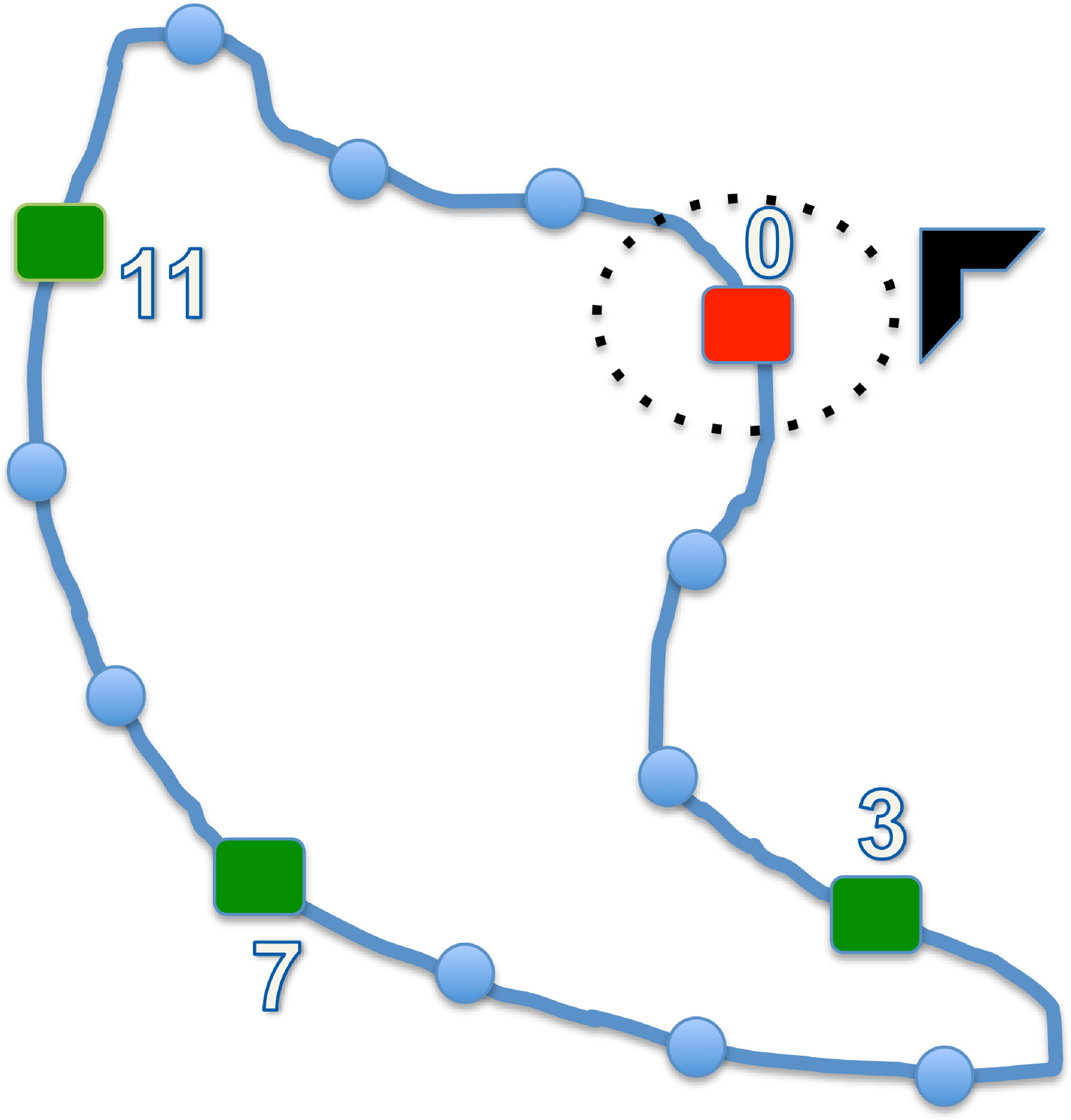}}
	{Perimeter patrol scenario with UAV loitering at alert station.\label{fig:patrol}}
	{}
\end{figure}

The perimeter patrol problem arose from the Cooperative Operations in Urban Terrain (COUNTER) project at AFRL \citep{Gross}. In this problem, there is a perimeter which must be monitored by a collection of UAVs (we will consider only one UAV here). Along the perimeter, there are $m$ alert stations equipped with Unattended Ground Sensors (UGSs) which detect intrusions or incursions into the perimeter. For the sake of simplicity, we assume that incursions into the perimeter can only occur at the stations.
An incursion could be a nuisance (false alarm) or a real threat. The UGS raise an alarm or an alert whenever there is an incursion.
The camera equipped UAV responds to an alert by flying to the alert site and loitering there, while a remotely located operator steers the gimballed camera looking for the source of the alarm. Here the operator serves the role of a classifier or a sensor, i.e., the operator must determine, from the video information, whether the intrusion is a nuisance or a threat. For details on the perimeter alert patrol problem and the variants thereof, we refer the reader to the authors' prior work \citep{Chandler,Darbha,Krishna10,KrishnaACC}. Figure~\ref{fig:patrol} shows a typical scenario, where there are $4$ alert stations with the UAV at a station (location $0$) with an alert.
The decision problem we solve is the following: Given that the arrival process of the alerts is Poisson with known arrival rate, what is the optimal time a UAV should spend at a station before resuming its patrol? 
We associate an information gain with a UAV loitering and servicing an alert and we model this gain as a monotonically increasing function of the loiter/dwell time $d$. 

\subsection{Problem Statement}
\label{sec:probstat}
\begin{figure}
\psfragscanon%
%
\psfrag{s01}[t][t]{\color[rgb]{0,0,0}\setlength{\tabcolsep}{0pt}\begin{tabular}{c}Dwell Time ($d$) \end{tabular}}%
\psfrag{s02}[b][b]{\color[rgb]{0,0,0}\setlength{\tabcolsep}{0pt}\begin{tabular}{c}Information Gain ($\mathcal I$)\end{tabular}}%
\psfrag{s03}[b][b]{\color[rgb]{0,0,0}\setlength{\tabcolsep}{0pt}\begin{tabular}{c}Value of Information gained versus Dwell time\end{tabular}}%
%
\psfrag{x01}[t][t]{0}%
\psfrag{x02}[t][t]{0.1}%
\psfrag{x03}[t][t]{0.2}%
\psfrag{x04}[t][t]{0.3}%
\psfrag{x05}[t][t]{0.4}%
\psfrag{x06}[t][t]{0.5}%
\psfrag{x07}[t][t]{0.6}%
\psfrag{x08}[t][t]{0.7}%
\psfrag{x09}[t][t]{0.8}%
\psfrag{x10}[t][t]{0.9}%
\psfrag{x11}[t][t]{1}%
\psfrag{x12}[t][t]{0}%
\psfrag{x13}[t][t]{1}%
\psfrag{x14}[t][t]{2}%
\psfrag{x15}[t][t]{3}%
\psfrag{x16}[t][t]{4}%
\psfrag{x17}[t][t]{5}%
\psfrag{x18}[t][t]{6}%
%
\psfrag{v01}[r][r]{0}%
\psfrag{v02}[r][r]{0.1}%
\psfrag{v03}[r][r]{0.2}%
\psfrag{v04}[r][r]{0.3}%
\psfrag{v05}[r][r]{0.4}%
\psfrag{v06}[r][r]{0.5}%
\psfrag{v07}[r][r]{0.6}%
\psfrag{v08}[r][r]{0.7}%
\psfrag{v09}[r][r]{0.8}%
\psfrag{v10}[r][r]{0.9}%
\psfrag{v11}[r][r]{1}%
\psfrag{v12}[r][r]{0}%
\psfrag{v13}[r][r]{0.005}%
\psfrag{v14}[r][r]{0.01}%
\psfrag{v15}[r][r]{0.015}%
\psfrag{v16}[r][r]{0.02}%
\psfrag{v17}[r][r]{0.025}%
\psfrag{v18}[r][r]{0.03}%
%


\FIGURE
	{\includegraphics*[width=4in]{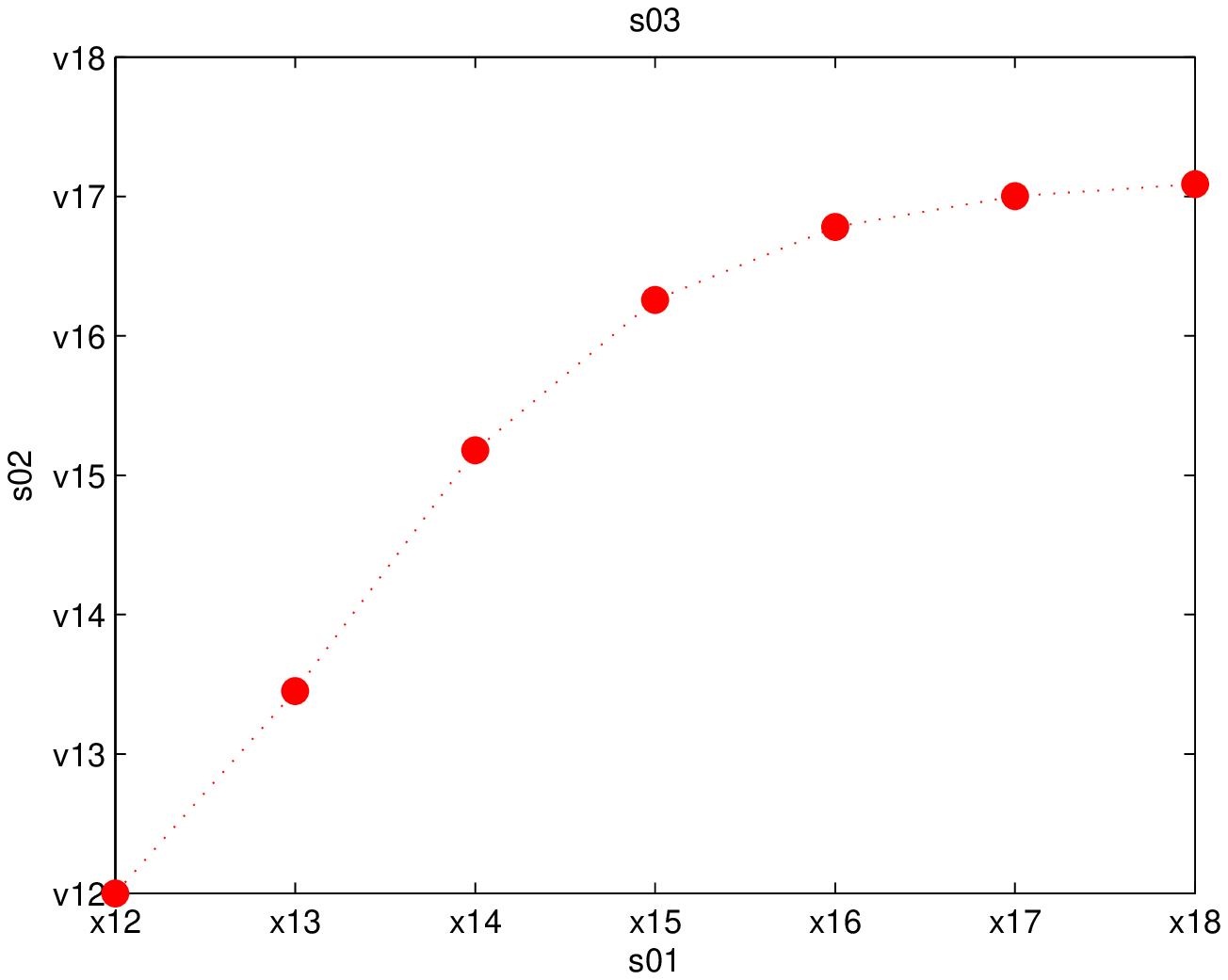}}
	{Value of Information gained vs dwell time.\label{fig:info}}
	{}
\end{figure}

The patrolled perimeter is a simple closed curve with $N (\geq m)$ nodes which are (spatially) uniformly separated, of which $m$ correspond to the alert stations. Let the $m$ distinct station locations be elements of the set $\Omega\subset\{0,\ldots,N-1\}$. A typical scenario shown in Figure~\ref{fig:patrol} has $15$ nodes, of which, nodes $\{0,3,7,11\}$ correspond to the UGS. Here, station locations $3$, $7$ and $11$ have no alerts, and station location $0$ has an alert being serviced by the loitering UAV.
At time instant $t$, let $\ell(t)$ be the position of the UAV on the perimeter ($\ell\in\{0,\ldots,N-1\}$), $d(t)$ be the dwell time (number of loiters completed if at an alert site) and $\tau_j(t)$ be the delay in servicing an alert at location $j\in\Omega$. 
Let 
$y_j(t)$ be a binary, but random, variable indicating the arrival of an alert at location $j\in\Omega$. We will assume that the statistics associated with the random variable $y_j(t)$ are known and that $y_j; j\in\Omega$ are independent. We model the arrival of alerts as follows:
There is a single queue with a Poisson arrival stream of alerts at a rate of $\alpha$ alerts per unit time. After an alert is queued up, we assume it shows up arbitrarily at any one of the $m$ stations (assuming choice of station is a uniformly distributed random variable). For this reason, only one alert can arrive at one of the $m$ stations at any instant of time. Hence, there are $m+1$ possibilities for the value of the vector of alerts $y(t) = [y_1(t) \; y_2(t) \; \cdots y_m(t)]$, with the first one being that there is no alert at any station and the other $m$
correspond to an alert at each of the $m$ stations. The control decisions are indicated by the variable $u$.  If $u =1$, then the UAV continues in the same direction as before; if $u =-1$, then the UAV reverses its direction of travel and if $u = 0 $, the UAV dwells at the current alert station. 
We will assume that a UAV advances by one node in unit time if $u\neq 0$. We also assume that the time to complete one loiter is also the unit time.
We denote the UAV's direction of travel by $\omega$, where $\omega=1$ and $\omega=-1$ indicate the clockwise and counter-clockwise directions respectively. One may write the state update equations for the system as follows:
\begin{eqnarray}
\label{eq:steq}
 \ell(t+1) &=& [\ell(t) + \omega(t)u(t)] \mod{N} , \nonumber\\
 \omega(t+1) &=& \omega(t) u(t)+\delta(u(t)), \nonumber\\
 d(t+1) &=& (d(t) + 1) \delta(u(t)),\\
\tau_j(t+1) &=& (\tau_j(t)+1)\left\{(1-\delta(\ell(t)-j)\delta(u(t))\right\}\max\left\{
\sigma(\tau_j(t)), y_j(t)\right\}, \quad \forall j\in\Omega,
\nonumber\end{eqnarray}
where $\delta$ is the Kronecker delta function and $\sigma(\cdot)=1-\delta(\cdot)$. We denote the status of the alert at station location $j\in\Omega$ at time $t$ by ${\cal A}_j(t)$, i.e., 
\begin{equation}
\label{eq:alstatus}
{\cal A}_j(t)= \left\{\begin{array}{l} 0, \mbox{ if } \tau_j (t)= 0 \\
1, \mbox{ otherwise} \end{array}\right. , \forall j\in\Omega.
\end{equation}
Also, we have the constraints:
$u(t) = 0$ only if  $\ell(t)\in\Omega$ and
$d(t) \leq D$.
If $d(t)=D$, then $u(t)\neq 0$ i.e., the UAV is forced to leave the station if it has already completed the maximum (allowed) number of dwell orbits. Combining the different components in (\ref{eq:steq}), 
we express the evolution equations compactly as:
\begin{equation}
\textbf{x}(t+1)=f(\textbf{x}(t),u(t),y(t)),\nonumber
\end{equation}
where,
$\mathbf x(t)$ is the system state at time $t$ with components $\ell(t)$, $\omega(t)$,$d(t)$ and $\tau_j(t)$, $\forall j\in\Omega$. 
Let us denote the $m+1$ possible values that $y(t)$ can take by the row vector $Y_l$ where,
\begin{eqnarray}
\label{eq:ydefn}
Y_0 = \left[\begin{array}{cccc}0 & 0 & \ldots & 0 \end{array} \right],\quad
Y_1 = \left[\begin{array}{cccc}1 & 0 & \ldots & 0 \end{array}\right],\quad
\ldots\quad\mbox{and}\quad
Y_{m} = \left[\begin{array}{cccc}0 & \ldots & 0 & 1 \end{array}\right].
\end{eqnarray}
Given a Poisson arrival stream of alerts at the rate of $\alpha$ alerts per unit time, the probability that there is no alert in unit time interval is $p=e^{-\alpha}$ and hence, the probability that $y(t)$ takes any one of the $m+1$ possible values in (\ref{eq:ydefn})
is given by, \begin{eqnarray}
\label{eq:prob}
p_l := Prob\{y(t)=Y_l\} = \left\{\begin{array}{ll}
                    p, \quad & l=0, \\
                    \frac{(1-p)}{m}, \quad & l=1,\ldots, m.
                    \end{array}\right.
                    \end{eqnarray}
To be consistent with the notation introduced earlier (in Sec~\ref{sec:sdp}), we shall use $\mathcal S$ to denote the set of all system states and use $x\in\{1,\ldots,|\mathcal S|\}$ to denote a particular state.                    
Our objective is to find a suitable policy that simultaneously minimizes the service delay and maximizes the information gained upon loitering. The information gain, $\mathcal I$, which is based on an operator error model (see Appendix~\ref{sec:opmodel}), is plotted as a function of dwell time in fig.~\ref{fig:info}. We model the one-step payoff/ reward function as follows:
\begin{equation}
\label{eq:patrwd}
R_u(x)=\left[\mathcal I(d_x+1)-  \mathcal I(d_x)\right]\delta(u)-\rho \; \max\{\bar\tau_x, \Gamma\},\quad x=1,\ldots,|\mathcal S|,
\end{equation}
where $d_x$ is the dwell associated with state $x$ and $\bar\tau_x=\max_{j\in\Omega} \tau_{j,x}$ is the worst service delay (among all stations) associated with state $x$. The parameter $\Gamma(>>0)$ is a judiciously chosen maximum penalty. 
%
%
%

The positive parameter $\rho$ is a constant weighing the incremental information gained upon loitering once more at the current location against the delay in servicing alerts at other stations. 
From the state definition, we can compute the total number of states in the MDP to be, 
\begin{equation}
\label{eq:numStates}
|\mathcal S|=2 \times N \times (\Gamma+1)^m + D\times m \times (\Gamma+1)^{m-1},
\end{equation}
where, the factor $2$ comes from the UAV being bi-directional. 
For the loiter states, directionality is irrelevant and hence when $d\geq1$, we reset $\omega$ to be $1$. Note that, in lieu of the reward function defintion (\ref{eq:patrwd}), we do not keep track of delays beyond $\Gamma$ and hence the state-space $\mathcal S$ only includes states $x$ with  $\tau_i\leq\Gamma,\forall i\in\Omega$ and so, is finite. We immediately see that the problem size is an $m^{th}$ order polynomial in $\Gamma$ and hence solving for the optimal value function and policy using exact dynamic programming (DP) methods are rendered intractable for practical values of $\Gamma$ and $m$.  Hence, we employ the restricted LP approach developed earlier to compute approximate value functions; from which we compute the corresponding greedy sub-optimal policy. In the next section, we exploit the structure in the perimeter patrol problem to simplify the $RLP$ and $NLP$ formulations and show that both collapse to exact LPs corresponding to MDPs defined on the $M$ partitions.

\subsection{Structure associated with the Perimeter Patrol Problem}
\label{sec:PatrolStruc}
In the perimeter patrol problem considered herein, we see that, by definition (\ref{eq:patrwd}), the reward function $R_u({x})$ is bounded. Consequently the optimal value function is bounded.
To explain the inherent structure in the reward, consider a station where an alert is being serviced by a UAV. The information gained by the UAV about the alert is only a function of the service delay at the station and the amount of time the UAV dwells at the station servicing the alert. There is a natural partitioning of states; where no matter what the delays are at the other stations, the reward is the same, as long as the maximum delay and the dwell time of the UAV at the station are the same. 
So, we aggregate all the states which have the same values for $\ell,\; \omega,\; d,\; {\cal A}_j,\;\forall j\in \Omega$ and $\bar \tau=\max_{j\in\Omega} \tau_j$, into one partition. 
As a result of aggregation, the number of partitions can be shown to be, 
\begin{equation}
\label{eq:numPart}
M=2\times N + 2\times N \times (2^ {m}-1) \times \Gamma + m\times D + m\times D\times (2^{m-1}-1) \times \Gamma ,
\end{equation}
which is linear in $\Gamma$ and hence considerably smaller than the total number of states (\ref{eq:numStates}). 

We introduce the following notation, that will be used hereafter: Let $\ell_x, d_x, \omega_x, \tau_{j,x}$ and $\mathcal A_{j, x}$ represent respectively, the location, dwell, direction of UAV's motion and the service delay and alert status at station location $j\in\Omega$ corresponding to some state $x\in\{1,\ldots,|\mathcal S|\}$. Also, we will use $\ell(i),d(i),\omega(i)$, $\bar\tau(i)$ and $\mathcal A_j(i)$ 
to denote the location, dwell, direction, maximum delay, and the alert status at station location $j\in\Omega$ that correspond to some partition index $i\in\{1,\ldots,M\}$. 
We will also denote by ${x}(t;{x}_0,\mathbf u_t,\mathbf y_t)$ the state at time $t>0$; if the initial state at $t=0$ is ${x}_0$ and the sequence of inputs, $\mathbf u_t=\{u(0),u(1),\ldots,u(t-1)\}$ and disturbances, $\mathbf y_t=\{y(0),y(1),\ldots,y(t-1)\}$.
We also introduce a partial ordering of the states according to: $x \geq y$ iff $\ell_x = \ell_y,\; d_x= d_y, \; \omega_x = \omega_y$ and $\tau_{j,x} \geq \tau_{j,y},$  $\forall j\in\Omega$. By the same token, we also partially order partitions, $\mathcal S_i \geq \mathcal S_j$ iff for every ${z} \in \mathcal S_j$, there exists an ${x} \in \mathcal S_i$ such that ${x} \geq{z}$.
Recall that $\mathcal{T}(i,u)$ is the set of all distinct $(m+1)$ tuples of partition indices, that the system can transition to, from partition $\mathcal S_i$ under control action $u$. For the sake of notational simplicity, we denote the $l^{th}$ component of any tuple $k \in {\cal T}(i,u)$ by $k_{l-1}$ and the cardinality of the set ${\cal T}(i,u)$ by $|{\cal T}(i,u)|$. Also we define the partitions to be of two types:
a partition ${\mathcal S_i}$ is of type 1 and we write $i\in\mathcal P_1$ if 
$
\ell(i) \in \Omega, \; d(i) = 0, \; \mathcal A_{\ell(i)}(i)=1, \;\mbox{and }
\mathcal A_{j}(i)=1, \mbox{ for some } j\in\Omega,j\neq \ell(i),\;$
i.e., the UAV is at a station with an alert, the dwell time is zero and also there is an alert at some other station. 
Else it is of type 2 and we write $i\in\mathcal P_2$ . Given this definition, we have the following important result, that we will make use of, in the remainder of the paper.
\begin{lemma} 
\label{lem:mult}
The cardinality of $\mathcal T(i,u)$ is given by:
\[|\mathcal T(i,u)|=\left\{\begin{array}{l}\bar\tau(i),\;i\in\mathcal P_1\mbox{ and } u=0,\\
1,\mbox{ otherwise. }\end{array}\right.\]
\end{lemma}
\proof{Proof of Lemma~\ref{lem:mult}.} 
First we consider partition index $i$ of type 1 and control input $u=0$. Since the UAV has decided to loiter at the current station i.e., $\ell(i)\in\Omega$, the service delay at that station, $\tau_{\ell(i)}$ will be reset to zero in the next time step. Hence the future state (and partition) maximum delay will be determined by the highest of the service delays, say $\tilde\tau$, among the other stations with alerts (at least one such station exists since partition $i$ is of type 1). So $\forall j\in\{1,\ldots,\bar\tau(i)\},\;\exists\; x_j\in\mathcal S_i$ such that $\tilde\tau_{x_j}=j$. The corresponding tuple of future partition indices $z^{i,0}_{x_j} = (\bar{f}(x_j,u,Y_0),\bar{f}(x_j,u,Y_1),\hdots,\bar{f}(x_j,u,Y_m))$ will have maximum delay $j+1$ and so $\mathcal T(i,0)=\cup_{j=1}^{\bar\tau(i)} \{z^{i,0}_{x_j}\}\Rightarrow|\mathcal T(i,0)|=\bar\tau(i)$. For all other control choices, $u\neq0$, all the states $x\in\mathcal S_i$ will transition to future states with the same maximum delay $\bar\tau(i)+1$. So, for $u\neq0$, $\mathcal T(i,u)$ is a singleton set and hence $|\mathcal T(i,u)|=1$.
For partition indices $j$ of type 2 with $\bar\tau(j)>0$, all the states $x\in\mathcal S_j$ will transition to future states with the same maximum delay $\bar\tau(j)+1$ and so $|\mathcal T(j,u)|=1,\;\forall u$. If $\bar\tau(j)=0$, then the partition $\mathcal S_j$ is a singleton set as per the aggregation scheme (see Sec~\ref{sec:PatrolStruc}) and hence $|\mathcal T(j,u)|=1,\;\forall u$.
\Halmos 
\endproof

\begin{theorem} 
\label{th:NLPeqLP}
For the perimeter patrol problem, the $NLP$ (\ref{eq:NLPcon}) reduces to the following LP.
\begin{eqnarray}
\label{eq:NLPeqLP}
LBLP &:=& \min \bar c^T w, \quad\mbox{subject to}\nonumber\\
 w(i)&\geq& r_u(i) + \lambda \sum_{l=0}^{m} p_l w(k_l), \; \forall u,\; i=1,\ldots,M,
 \end{eqnarray} 
where the tuple $k\in\mathcal T(i,u)$, if $|\mathcal T(i,u)|=1$, else $k=k^*$, where  $k^*\in\mathcal T(i,u)$ is the tuple of partition indices such that $\bar\tau(k^*_l)=\bar \tau(i) +1,\;l=0,\ldots,m$.  
Furthermore, the optimal solution, $w^*$ is dominated by every feasible $w$  for the $NLP$ and, in particular, it is a lower bound to the optimal value function i.e., for all $i=1,\ldots,M$, one has  $w^*(i) \leq \min_{{x} \in {\cal S}_i}V^*({x})$.
\end{theorem}
Before proceeding further, we make two key claims that are essential for the proof of Theorem~\ref{th:NLPeqLP}. The justification for the claims have been provided in the Appendix.
\begin{claim}
\label{cl:monoSt}
If $x_1 \geq x_2$, then for the same sequence of inputs $\mathbf u_t$ and disturbances $\mathbf y_t$, the system state evolves in such a way that ${x}(t; x_1,\mathbf u_t,\mathbf y_t) \geq {x}(t; x_2,\mathbf u_t,\mathbf y_t) $ for every $t > 0$.
\end{claim}
\begin{claim}
\label{cl:monoVal}
If $x_1 \geq x_2$, then $V^*(x_1) \leq V^*(x_2)$. Furthermore, if $\mathcal S_i\geq\mathcal S_j$, then $\min_{x \in {\cal S}_i} V^*(x) \leq \min_{{z} \in {\cal S}_j} V^*({z})$.
\end{claim}

\proof{Proof of Theorem~\ref{th:NLPeqLP}.} 
Recall the non-linear constraints (\ref{eq:NLPmot}) satisfied by $\bar w(i):=\min_{x\in\mathcal S_i}V^*(x)$ that motivated the $NLP$ formulation:
\begin{eqnarray}
\label{eq:minValfn1}
\bar w(i)&\geq& \min_{x\in\mathcal S_i}\left\{R_u(x) + \lambda \sum_{l=0}^{m}p_l \bar w(\bar f(x,u,Y_l))\right\}, \quad\forall u,\quad i=1,\ldots,M,
\end{eqnarray}
which, given the definition of ${\cal T}(i,u)$, can be written in the following equivalent form:
\begin{equation}
\label{eq:NLPcon1}
\bar w(i)\geq r_u(i) + \lambda\min_{k \in {\cal T}(i,u)} \sum_{l=0}^{m} p_l \bar w(k_l), \quad\forall u,\quad i=1,\ldots,M,
 \end{equation}
where $r_u(i)$ is the reward associated with partition index $i$, and given the partitioning scheme, satisfies $R_u(x) = r_u(i), \forall x\in \mathcal S_i$.
Given the structure in the perimeter patrol problem, we will show that the above 
(\ref{eq:NLPcon1}) will collapse to a single linear inequality constraint for every partition index $i$ and control $u$. 
Let us focus our attention on partition index $i$ of type 1 and control action $u=0$. For this choice, the cardinality of ${\cal T}(i, 0)$ is $\bar \tau(i)$ as per Lemma~\ref{lem:mult}. 
Indeed $\exists\;\bar x\in\mathcal S_i$ such that the corresponding tuple of future partition indices $k^* = (\bar{f}(\bar x,0,Y_0),\bar{f}(\bar x,0,Y_1),\hdots,\bar{f}(\bar x,0,Y_m))$ has the highest possible maximum delay, i.e., $\bar\tau(k^*_l)=\bar\tau(i)+1, l=0,\ldots,m$. Since $k^*_l\geq k_l,\;l=0,\ldots,m,\;\forall k\in\mathcal T(i,u)$, we have from Claim~\ref{cl:monoVal} that, $\bar w(k^*_l)\leq \bar w(k_l),\;l=0.\ldots,m,\;\forall k\in\mathcal T(i,u)$. 
So, the non-linear inequality corresponding to partition index $i\in\mathcal P_1$ and control $u=0$ becomes:
\begin{eqnarray}
\label{eq:NLPcon2}
\bar w(i) &\geq& r_{0}(i) + \lambda \sum_{l=0}^{m} p_l \bar w({k_l^*}).
 \end{eqnarray}
If $u\neq 0$, then $|{\cal T}(i, u)|=1$. So there exists exactly one tuple ${\underbar k}$ in ${\cal T}(i, u)$ and hence, the non-linear constraint (\ref{eq:NLPcon1}) reduces to the linear inequality:
\begin{equation}
\label{eq:lincon}
\bar w(i) \geq r_u(i) + \lambda \sum_{l=0}^{m} p_l \bar w({{\underbar k}_l}).
\end{equation}
For partition indices $j$ of type 2, $|{\cal T}(j,u)|=1,\;\forall u$. So, as before, the non-linear inequality (\ref{eq:NLPcon1}) collapses to the linear inequality (\ref{eq:lincon}). 

In summary, we have the following: regardless of which partition one considers, the corresponding non-linear constraint in $NLP$ collapses to a linear constraint and hence, $NLP$ for the perimeter patrol problem collapses to the following LP:
\begin{eqnarray}
\label{eq:NLPeqLP1}
LBLP &:=& \min \bar c^T w, \quad\mbox{subject to}\nonumber\quad\\
 w(i)&\geq& r_u(i) + \lambda \sum_{l=0}^{m} p_l w(k_l), \quad \forall u,\; i=1,\ldots,M,
 \end{eqnarray} 
where the tuple $k\in\mathcal T(i,u)$, if $|\mathcal T(i,u)|=1$, else $k=k^*$, where  $k^*\in\mathcal T(i,u)$ is the tuple of partition indices such that $\bar\tau(k^*_l)=\bar \tau(i) +1,\;l=0,\ldots,m$. 
 
To prove the second part of the Theorem, 
we observe that $LBLP$ defined above is the exact LP corresponding to a reduced order MDP defined on the $M$ partitions. 
Hence, we readily have from Lemmas~\ref{lem:autoUB1} and \ref{lem:LPopt1} that the optimal solution $w^*$ lower bounds every feasible solution including $\bar w$ and hence,
$w^*(i) \leq \bar w(i) = \min_{{x} \in {\cal S}_i} V^*({x}) \leq V^*({y}), \; \forall{y} \in {\cal S}_i, \;  i=1,\ldots,M.$
\Halmos 
\endproof
So, for the perimeter patrol problem, one can compute a lower bound for the optimal value function efficiently by solving $LBLP$. The next logical question is whether the upper bound formulation, $RLP$ (\ref{eq:RLP}), also simplifies, given the structure in the problem. It turns out that this is indeed the case, as can be seen from the following theorem.
\begin{theorem} 
\label{th:RLPeqLP}
For the perimeter patrol problem, the $RLP$ (\ref{eq:RLP}) reduces to the following LP. 
\begin{eqnarray}
\label{eq:RLPeqLP}
UBLP &:=& \min \bar c^T w, \quad\mbox{subject to}\nonumber\\
 w(i)&\geq& r_u(i) + \lambda \sum_{l=0}^{m} p_l w(k_l), \quad \forall u,\quad i=1,\ldots,M,
 \end{eqnarray} 
where the tuple $k\in\mathcal T(i,u)$, if $|\mathcal T(i,u)|=1$, else $k=k^*$, where  $k^*\in\mathcal T(i,u)$ is the tuple of partition indices such that $\bar\tau(k^*_l)=2,\;l=0,\ldots,m$. 
\end{theorem}
\proof{Proof of Theorem~\ref{th:RLPeqLP}.} 
Given the partitioning scheme, one can rewrite the Bellman inequalities (\ref{eq:bellineq}) as follows: for each $i=1,\ldots,M$,
\begin{equation}
\label{eq:bellineq1}
V^*({x}) \ge {r_u}(i)+\lambda \sum_{l=0}^{m}{p_lV^*(f({x},u,Y_l))},\; \forall u,\;\forall x\in\mathcal S_i.
\end{equation}
With the restriction that $V(x)=v(i),\forall x\in\mathcal S_i$, we get the following constraint for $RLP$ (\ref{eq:RLP}),
\begin{equation}
\label{eq:RLPcon1}
 v(i)\geq r_u(i) + \lambda \sum_{l=0}^{m} p_l v(k_l), \; \forall k \in {\cal T}(i,u),\forall u,\; i=1,\ldots,M.
 \end{equation}
%
For partition index $i\in\mathcal P_1$, $\exists\;\bar x\in\mathcal S_i$ that transitions to future states with the least possible maximum delay, $2$. Hence $f(\bar x,0,Y_l)\leq f(x,0,Y_l),\;l=0,\ldots,m,\;\forall x\in\mathcal S_i$ and so from Claim~\ref{cl:monoVal} we have, $V^*(f(\bar x,0,Y_l))\geq V^*(f(x,0,Y_l)),\;l=0,\ldots,m,\;\forall x\in\mathcal S_i$.
So, for $i\in\mathcal P_1$ and $u=0$, the inequalities (\ref{eq:bellineq1}) can be written as follows,
 \begin{eqnarray}
\label{eq:bellineq2}
 V^*(x)\geq {r_0}(i)+\lambda \sum_{l=0}^{m}{p_lV^*(f(\bar{{x}},0,Y_l))}\; \forall u,\;\forall x\in\mathcal S_i.
 \end{eqnarray}
The above implies that the $\bar\tau(i)$ constraints (\ref{eq:RLPcon1}) in $RLP$ can be replaced by the single constraint,
\begin{equation}
\label{eq:RLPcon2}
 v(i)\geq r_0(i) + \lambda \sum_{l=0}^{m} p_l v({k^*_l}),
 \end{equation}
where $k^* = (\bar{f}(\bar x,0,Y_0),\bar{f}(\bar x,0,Y_1),\hdots,\bar{f}(\bar x,0,Y_m))$ is the tuple of future partition indices (corresponding to $\bar x$) with the least possible maximum delay, i.e., $\bar\tau(k^*_l)= 2,\;l=1,\ldots,m$. For the other control choices, $u\neq 0$, there exists only one tuple  $\bar k$ in $\mathcal T(i,u)$ (since $|\mathcal T(i,u)|=1$) and hence the constraint (\ref{eq:RLPcon1}) is the single constraint,
\begin{equation}
\label{eq:RLPcon2}
 v(i)\geq r_u(i) + \lambda \sum_{l=0}^{m} p_l v({\bar k_l}),\;u\neq 0.
 \end{equation}
Similarly, for partitions $\mathcal S_j$ of type 2, $|\mathcal T(j,u)|=1,\;\forall u$, and so the constraint (\ref{eq:RLPcon1}) is the single constraint (\ref{eq:RLPcon2}).

In summary, we have the following: regardless of which partition index $i\in\{1,\ldots,M\}$ and control action $u$ are considered, the corresponding $|\mathcal T(i,u)|$ linear constraints in $RLP$ collapse to a single constraint and hence, $RLP$ for the perimeter patrol problem reduces to the following exact LP:
\begin{eqnarray}
\label{eq:RLPeqLP1}
UBLP &:=& \min \bar c^T w, \quad\mbox{subject to}\nonumber\\
 w(i)&\geq& r_u(i) + \lambda \sum_{l=0}^{m} p_l w(k_l), \quad \forall u,\quad i=1,\ldots,M,
 \end{eqnarray} 
where the tuple $k\in\mathcal T(i,u)$, if $|\mathcal T(i,u)|=1$, else $k=k^*$, where  $k^*\in\mathcal T(i,u)$ is the tuple of partition indices such that $\bar\tau(k^*_l)=2,\;l=0,\ldots,m$. 
\Halmos 
\endproof
In conclusion, we have two complementary LP formulations, $UBLP$ and $LBLP$ that can be used to efficiently compute upper bound and lower bound approximate value functions respectively, for the perimeter alert patrol problem. Note that the two formulations involve computing the optimal value functions for reduced order MDPs defined over the $M$ partitions and in that sense are computationally attractive (compared to solving the original problem) since $M<<|\mathcal S|$. In the following section, we will provide numerical results that corroborate the key claims made earlier regarding the structure in the perimeter alert patrol problem.
 
 \section{Numerical Results}
\label{sec:results}
We consider a perimeter with $N=15$ nodes of which node numbers $\{0,3,7,11\}$ are alert stations and a maximum allowed dwell of $D=5$ orbits. The other parameters were chosen to be: weighing factor, $\rho= .005$ and temporal discount factor, $\lambda=0.9$. Based on experience, we chose the alert arrival rate $\alpha=\frac{1}{60}$. This reflects a rather low arrival rate where we expect $2$ alerts to occur on average in the time taken by the UAV to complete an uninterrupted patrol around the perimeter. 
 We set the maximum delay time, that we keep track of, to be $\Gamma=15$; for which the total number of states comes out to be  $|\mathcal S|= 2,048,000$. Before venturing into the simulation, we first provide numerical results that corroborate the key Claim~\ref{cl:monoVal}, made earlier in the paper. For this, we solve for the optimal value function $V^*$. This is possible since the size of the example problem considered in this section is small and hence an exact solution can be obtained. 
\begin{figure}
\FIGURE
	{\includegraphics*[width=5in]{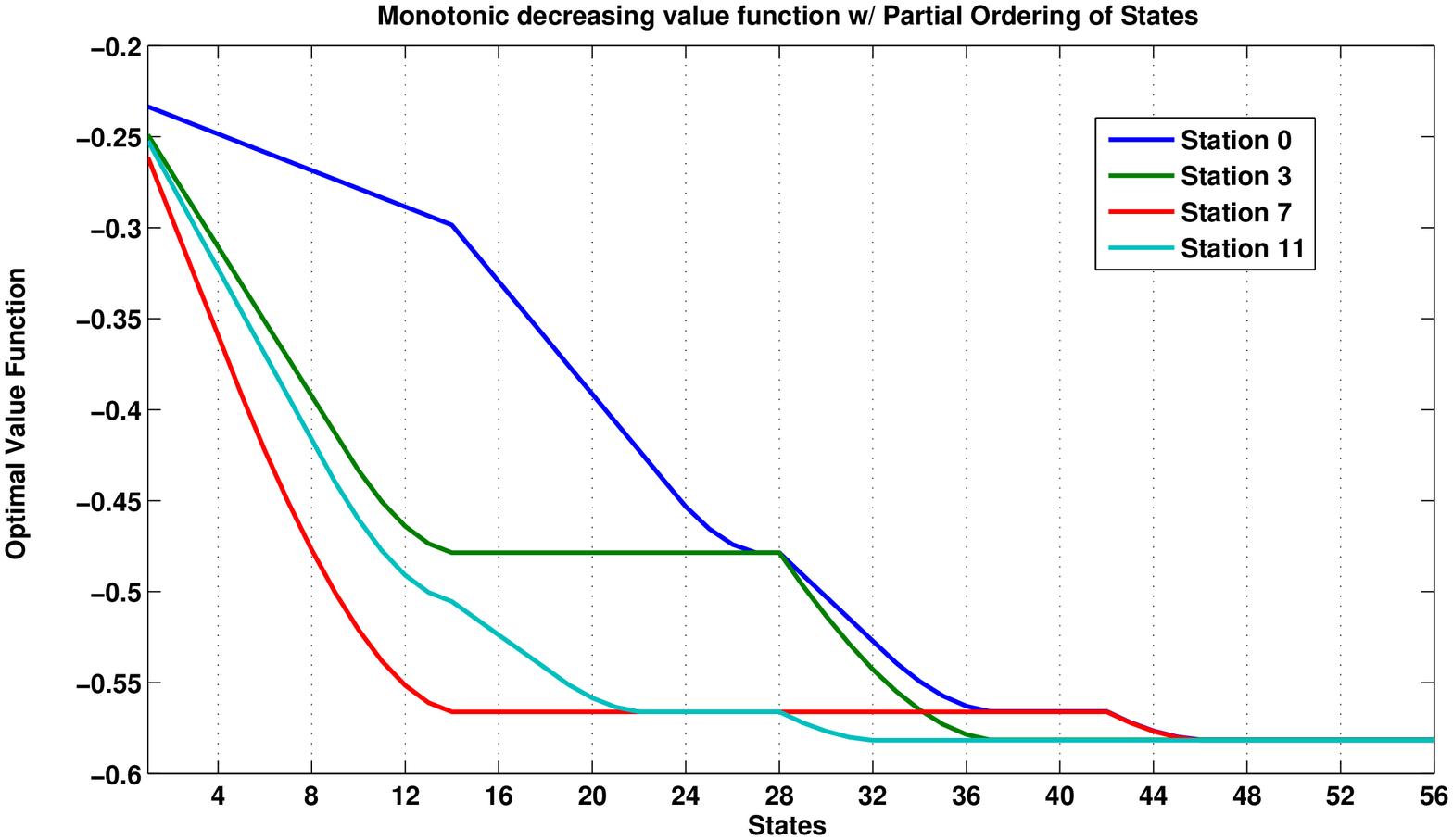}}
	{Monotonically decreasing value function corresponding to partially ordered states with increasing maximum delay.\label{fig:MonoVal}}
	{}
\end{figure}
In Figure~\ref{fig:MonoVal}, we show results supporting the claim that for partially ordered states $x_1\geq x_2$, the corresponding optimal value functions satisfy $V^*(x_1) \leq V^*(x_2)$. For this, we plot the optimal value function $V^*$ corresponding to states with alert status $\mathcal A_j=1,\forall j\in\Omega$ (all stations have alerts), dwell $d=0$, direction $\omega=1$ and the UAV located at one of the four station locations $\ell\in\Omega$. The partially ordered states represented in the X-axis are non-decreasing from left to right with maximum delay $\bar\tau$ varying from $2$ to $\Gamma$. The dotted grid lines in the plot separate the different partitions that the states fall into. 
\begin{figure}
\FIGURE
	{\includegraphics*[width=5in]{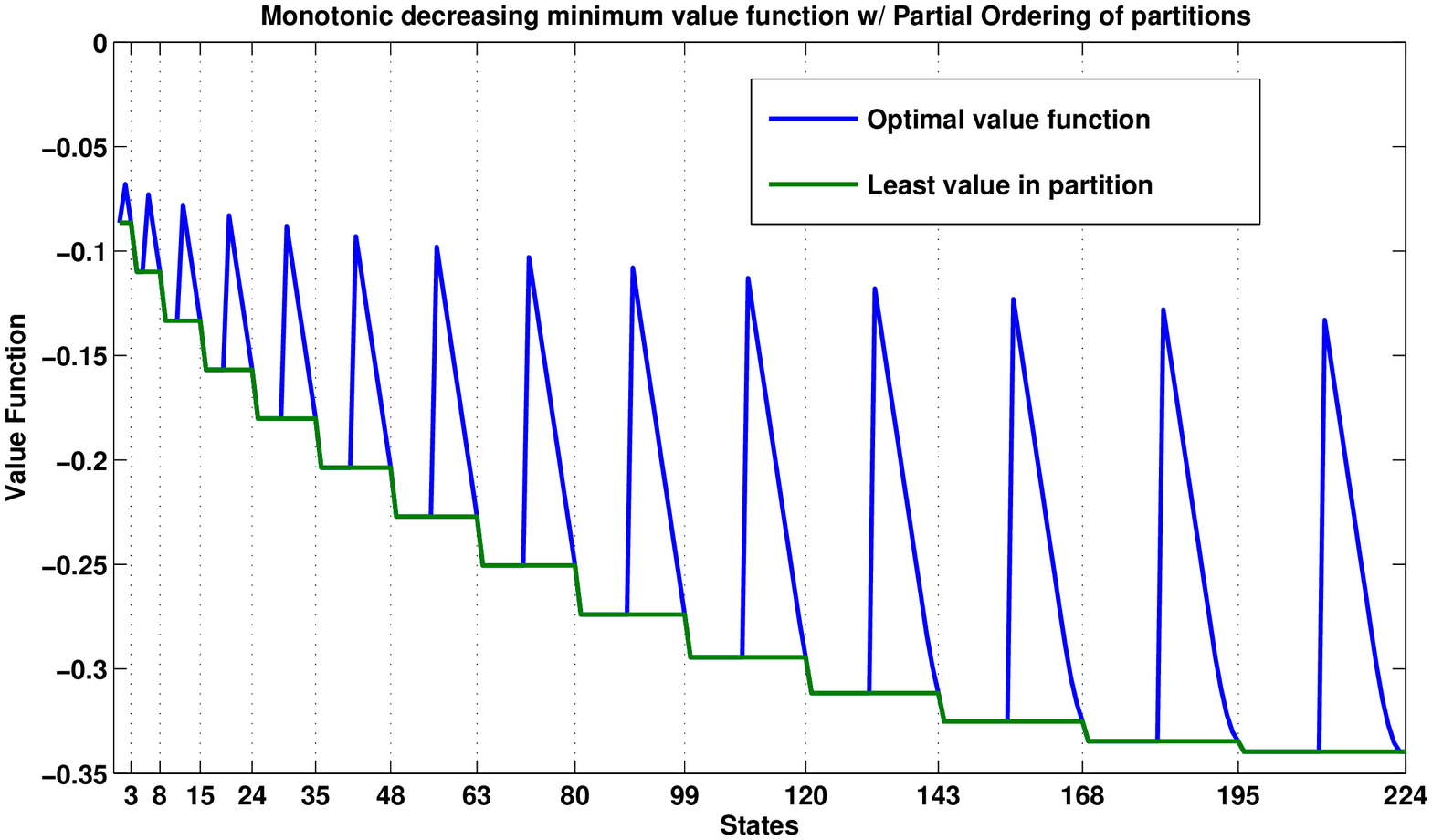}}
	{Monotonically decreasing least value function corresponding to partially ordered partitions with increasing maximum delay.\label{fig:MonoValPart}}
	{}
\end{figure}
In Figure~\ref{fig:MonoValPart}, we show results supporting the claim that for partially ordered partitions $\mathcal S_i\geq \mathcal S_j$, the corresponding optimal value functions satisfy $\min_{x\in\mathcal S_i}V^*(x) \leq \min_{y\in\mathcal S_j}V^*(y)$. For this, we plot the value functions corresponding to states with alert status $\mathcal A = 1 0 0 1$ (station locations $0$ and $11$ have alerts), dwell $d=0$, direction $\omega=1$ and $\ell=0$. The partially ordered partitions demarcated by the dotted grid lines in the X-axis are non-decreasing from left to right with maximum delay $\bar\tau$ varying from $2$ to $\Gamma$. Within each partition, we plot the value function associated with every state in the partition and also the least value function in the partition shown as the green line. One can easily see that the claim above is satisfied.

In the next section, we shall consider the same example problem and show that the proposed approximate methodology is effective. For this, we compute the approximate value functions via the restricted LP formulation and compare them with the optimal value function. In addition, we also compute the greedy sub-optimal policy corresponding to the approximate value function and compare it with the optimal policy in terms of the two performance metrics: alert service delay and information gained upon loitering.

 \subsection{Simulation Results}
 \label{sec:sim}
\begin{figure}
\FIGURE
	{\includegraphics*[width=5in]{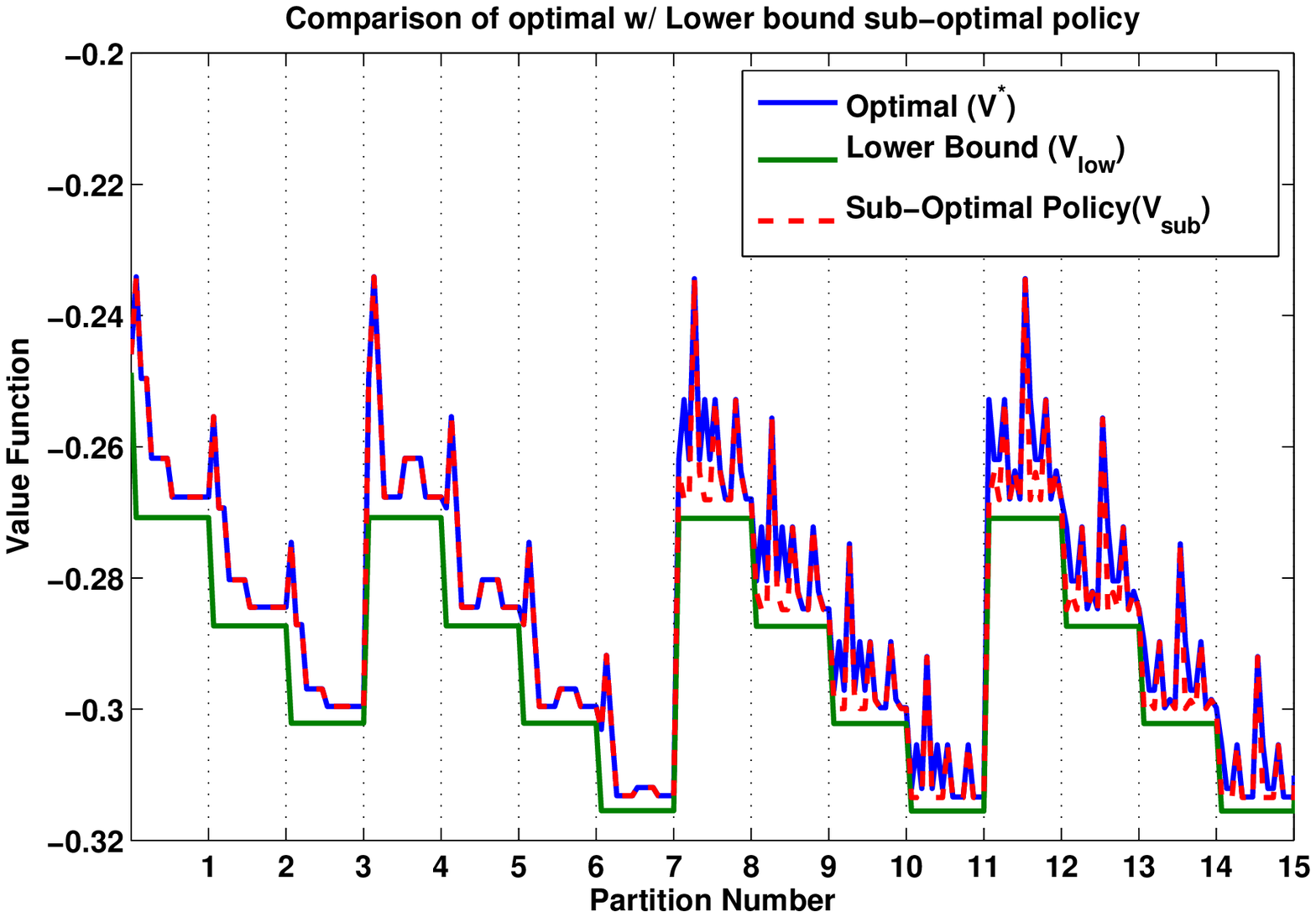}}
	{Comparison of value function corresponding to suboptimal policy $\pi_s$ with the optimal.\label{fig:BNDs_SUBOPT}}
	{}
\end{figure}

\begin{figure}
\FIGURE
	{\includegraphics*[width=5in]{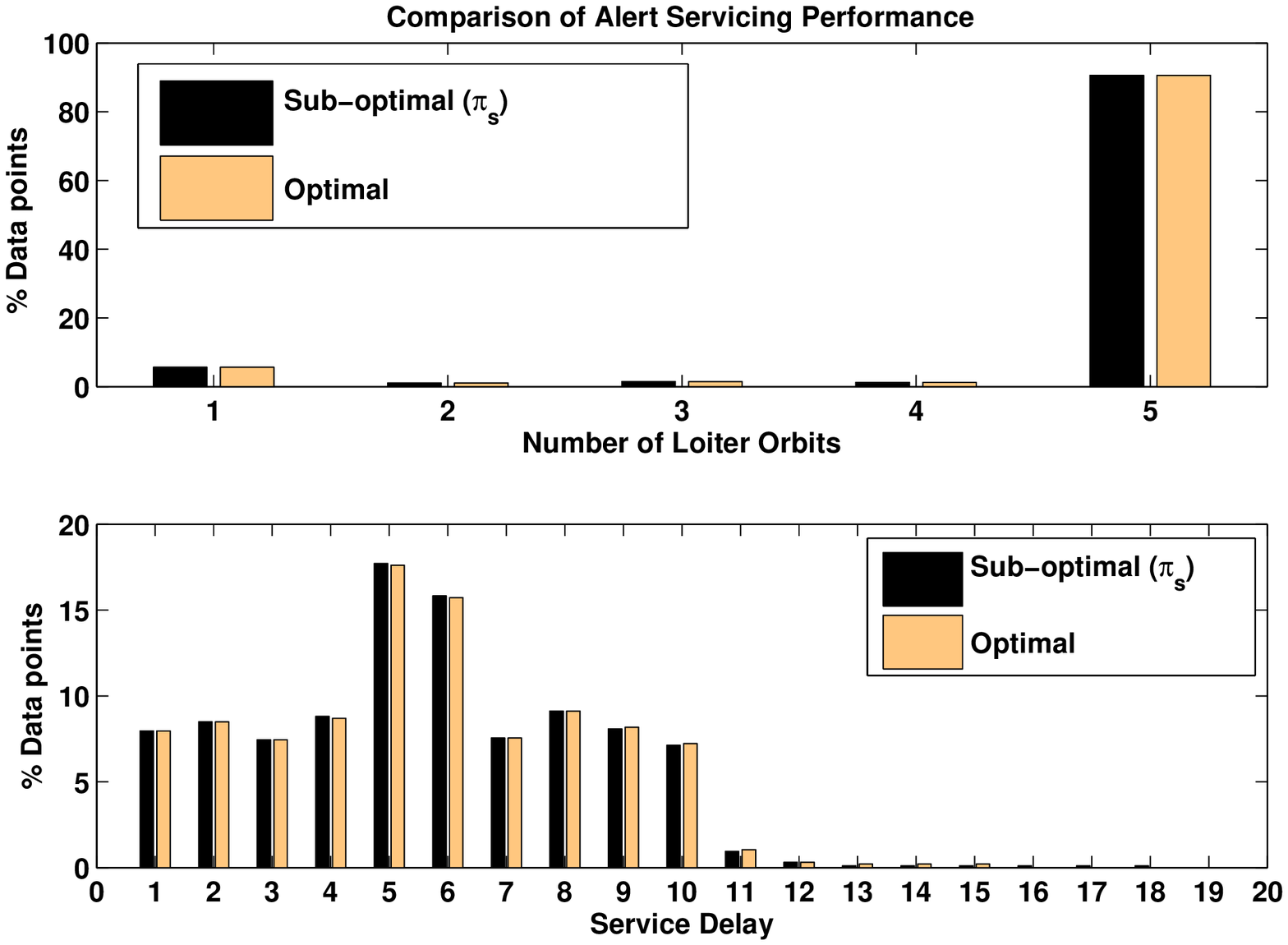}}
	{Comparison of service delay and number of loiters between optimal and sub-optimal policies.\label{fig:perfcomp}}
	{}
\end{figure}

We aggregate the states in the example problem based on the reward function (see section~\ref{sec:PatrolStruc} for details). This results in $M=8900$ partitions, which is considerably smaller than the original number of states, $|\mathcal S|$. We solve both the $UBLP$ and $LBLP$ formulations which give us the upper and lower bounds, $v^*$ and $w^*$ respectively, to the optimal value function $V^*$. Since we have the optimal value function for the example problem, we use it for comparison with the approximations. Note that for higher values of $m$ and $\Gamma$, the problem would essentially become intractable and one would not have access to the optimal value function. Nevertheless, one can compute $v^*$ and $w^*$ and the difference between the two would give an estimate of the quality of the approximation. 
\begin{figure}
\FIGURE
	{\includegraphics*[width=5in]{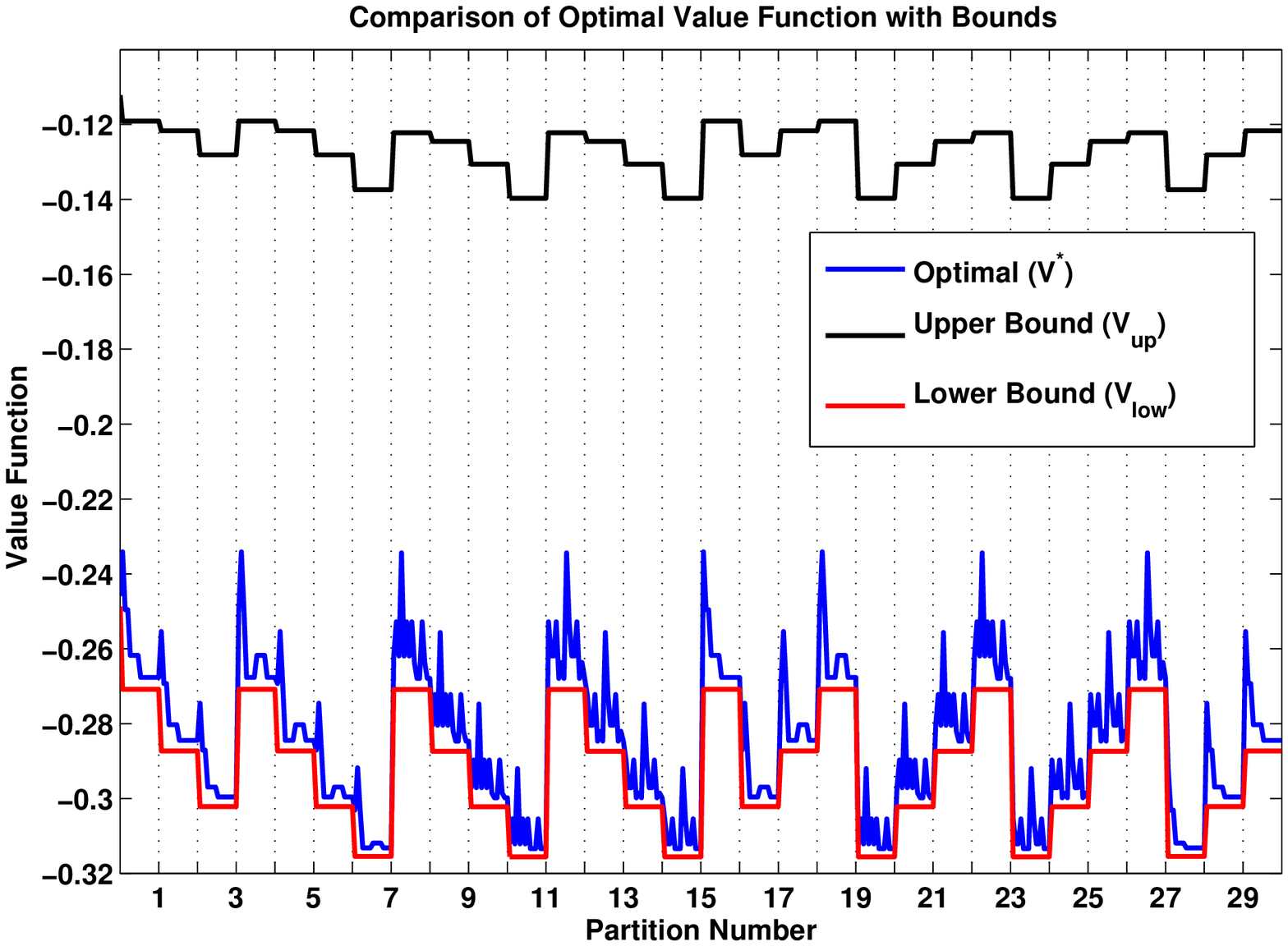}}
	{Comparison of approximate value functions with the optimal.\label{fig:BNDs_OPT}}
	{}
\end{figure}
We give a representative sample of the approximation results by choosing all the states in partitions corresponding to alert status $\mathcal A_j=1,\forall j\in\Omega$ (all stations have alerts) and maximum delay $\bar\tau=2$.
Figure~\ref{fig:BNDs_OPT} compares the optimal value function $V^*$ with the upper and lower bound approximate value functions, $ V_{up}=\Phi v^*$ and $ V_{low} = \Phi w^*$ for this subset of the state-space. The first $15$ partitions shown in the X-axis of Figure~\ref{fig:BNDs_OPT} i.e., partition numbers, $i=1,\ldots,15$, correspond to the clockwise states:
  \begin{equation}
  \label{eq:cw}
  \ell = i-1,\quad d = 0, \quad \omega=1, \quad \bar\tau=\max_{j\in \Omega}\tau_j=2, \quad
\mathcal  A_j=1, \;\forall j\in \Omega,
  \end{equation}
  and the last $15$ partitions shown in the X-axis i.e., partition numbers, $i=16,\ldots,30$, correspond to the counter-clockwise states:
\begin{equation}
  \label{eq:ccw}
  \ell = i-N-1,\quad d = 0, \quad \omega=-1, \quad \bar\tau=\max_{j\in \Omega}\tau_j=2, \quad
\mathcal  A_j=1,\; \forall j\in \Omega.
  \end{equation}
Interestingly, we notice immediately that the lower bound appears to be tighter than the upper bound. 
Recall that our objective is to obtain a good sub-optimal policy and so, we 
consider 
the policy that is \emph{greedy} with respect to $V_{low}$:
\begin{equation}
\label{eq:polimp}
\pi_{s}(x)=\arg\max_{u} \left\{  R_u(x) + \lambda  \sum_{l=0}^{m}{p_lV_{low}(f(x,u,Y_l))} \right\},\quad\forall x\in\{1,\ldots,|\mathcal S|\}.
\end{equation}
To assess the quality of the sub-optimal policy, we also compute the expected discounted payoff,  $V_{sub}$ that corresponds to the sub-optimal policy $\pi_{s}$, by solving the system of equations:
\begin{equation}
\label{eq:poleval}
(I-\lambda P_{\pi_{s}})V_{sub} = R_{\pi_{s}}.
\end{equation}
Since $V_{sub}$ corresponds to a sub-optimal policy and in lieu of the monotonicity property of the Bellman operator, the following inequalities hold:
\[V_{low}\leq V_{sub}\leq V^*\leq V_{up}.\]
In Figure~\ref{fig:BNDs_SUBOPT}, we compare $V_{sub}$ with the optimal value function $V^*$ for the clockwise states defined in (\ref{eq:cw}) and note that the approximation is quite good.
Finally, we compare the performance of the sub-optimal policy $\pi_s$ with that of the optimal strategy $\pi^*$ in terms of the two important metrics: service delay and information gain (measured via the dwell time). To collect the performance statistics, we ran Monte Carlo simulations with alerts generated from a Poisson arrival stream with rate $\alpha=\frac{1}{60}$ over a $60000$ time unit simulation window. Both the optimal and sub-optimal policies were tested against the same alert sequence. Figure~\ref{fig:perfcomp} shows histogram plots for the service delay (top plot) and the dwell time (bottom plot) for all serviced alerts in the simulation run. The corresponding mean and worst case service delays and the mean dwell time are also shown in Table~\ref{tab:perfcomp}. We see that there is hardly any difference in terms of either metric between the optimal and the sub-optimal policies. This substantiates the claim that the aggregation approach gives us a sub-optimal policy that performs almost as well as the optimal policy itself.
This is to be expected, given that the value functions corresponding to the optimal and sub-optimal policies are close to each other (see Figure~\ref{fig:BNDs_SUBOPT}). 
Since the false alarm rate $\alpha$ is fairly low, we see from the bottom plot of Figure~\ref{fig:perfcomp} that roughly $90\%$ of the alerts were cleared within ten time steps. Also from the top plot of Figure~\ref{fig:perfcomp}, we see that  maximum information was gained ($5$ loiters completed) on almost $90\%$ of the serviced alerts. 

\begin{table}
	\TABLE
	{Comparison of alert servicing performance between optimal and sub-optimal policies.	\label{tab:perfcomp}}
		{\begin{tabular}{|c|c|c|c|}
			\hline
			Policy & Mean number of loiters & Mean service delay & Worst service delay\\
			\hline
			$\pi^*$ & 4.7 & 5.6 & 15\\
			\hline
			$\pi_s$ & 4.7 & 5.6 & 18\\
			\hline
			\end{tabular}}
			{}
	\end{table}

\section{Conclusions}
\label{sec:con}
We have provided a state aggregation based restricted LP method to construct sub-optimal policies for stochastic DPs along with a bound for the deviation of such a policy from the optimum value function. 
As a key result, we have shown that the solution to the aggregation based LP is independent of the underlying cost function and we do so by demonstrating that the restricted LP is, in fact, the exact LP that corresponds to a lower dimensional MDP defined over the partitions. We also provide a novel non-linear program that can be used to compute a non-trivial lower bound to the optimal value function. In particular, for the perimeter patrol stochastic control problem, we have shown that both the upper and lower bound formulations simplify to exact LPs corresponding to some reduced order MDPs. To do so, we have exploited the partial ordering of the states that comes about because of the structure inherent in the reward function. It would be interesting to see if the simplification can be achieved for other problems that exhibit a similar structure. For the perimeter patrol problem, numerical results obtained via Monte Carlo simulations show that the sub-optimal policy obtained via the approximate value functions perform almost as well as the optimal policy. 
The literature suggests that, in general, the solution to a restricted LP depends on the underlying cost function; when the value function is parameterized by arbitrary basis functions. We have shown that, for the special case of hard aggregation, this is not true. Surely, there exist other basis functions with the same property and it would be useful to uncover the class of basis functions, for which the independence result holds.

\bibliographystyle{ormsv080} 
\bibliography{Ref_bnds} 


\ECSwitch

\ECDisclaimer

\ECHead{\centering Appendix to ``Bounding Procedures for Stochastic Dynamic Programs with Application to the Perimeter Alert Patrol Problem'' by Park et al.}

This appendix contains supplementary material to the paper and also lengthy proofs that were left out of the main document.

\section{Operator Error Model}
\label{sec:opmodel}
We  treat the operator as a sensor-in-the-loop automaton. The operator is not infallible and we account for that statistically in the optimization. To quantify the operator's performance, we consider two random variables: the variable $X$ that specifies whether the alert is a real threat (target $T$) or a nuisance (false target $FT$) and the operator decision $Z$ which specifies whether he determines the alert to be a real threat $Z_1$ or a nuisance $Z_2$. We stipulate that the a priori probability that an alert is a real target,
\begin{equation}
\label{eq:apriori}
Prob\{X=T\} = p << 1.
\end{equation}
We assume, based on experience, that $p=0.01$ in this work. The conditional probabilities which specify whether the operator correctly reported a threat and a nuisance are assumed to be functions of the dwell time, $d$:
\begin{eqnarray}
\label{eq:condprob}
 P_{TR}(d):=Prob\{Z=Z_1|X=T\} &=&a+b(1-e^{-\mu_{1}{d}}), \nonumber\\
 P_{FTR}(d):=Prob\{Z=Z_2|X=FT\} &=&c+g(1-e^{-\mu_{2}{d}}).
\end{eqnarray}
where the acronyms $TR$ and $FTR$ stand for \emph{Target Report} and \emph{False Target Report} respectively. 
The parameters $a$, $b$, $\mu_{1}$, $c$, $g$, $\mu_{2}$ characterize the ``confusion matrix'' and the performance of the operator as a sensor; for details on sensor performance modeling, see Sec 7.2 in \cite{Kish1}. The parameters satisfy the constraints:
\[0<a+b\leq 1,\quad 0<c+g\leq 1,\quad \mu_1\geq 0 \quad\mbox{and} \quad \mu_2\geq 0.\]
In this work, we chose $a=c=0.5$, $b=g=0.45$ and $\mu_{1}=\mu_{2}=1$. The choice $a=c=0.5$ correspond to an uninformed or unbiased operator, i.e., the operator cannot tell if the alert is a threat or a nuisance without having seen any video footage of the alert site. 
We wish to maximize the mutual information - derived along the lines of information theory \citep{Cover1} - between the random variables $X$ and $Z$ given by:
\begin{eqnarray}
\mathcal I(X;Z) &=& H(X)-H(X|Z) \nonumber\\
&=& \sum_{x,z}Prob\{X=x,Z=z\}\log\frac{Prob\{X=x,Z=z\}}{Prob\{X=x\}Prob\{Z=z\}},
\end{eqnarray}
where $H(X)$ is the entropy of $X$ and $H(X|Z)$ is the conditional entropy of $X$ given $Z$. Using Bayes' rule and the probabilities (\ref{eq:apriori}) and (\ref{eq:condprob}), one can show that the mutual information is a function of dwell time, $d$:
\begin{eqnarray}
  \mathcal I(d)&=&pP_{TR}\log\frac{P_{TR}}{pP_{TR}+(1-p)(1-P_{FTR})}\nonumber\\
&+&  p(1-P_{TR})\log\frac{1-P_{TR}}{p(1-P_{TR})+(1-p)P_{FTR}} \nonumber\\
&+& (1-p)(1-P_{FTR})\log\frac{1-P_{FTR}}{pP_{TR}+(1-p)(1-P_{FTR})} \nonumber\\
&+& (1-p)P_{FTR}\log\frac{P_{FTR}}{p(1-P_{TR})+(1-p)P_{FTR}},
\end{eqnarray}
since the conditional probabilities, $P_{TR}$ and $P_{FTR}$ are both functions of $d$ (\ref{eq:condprob}).
\section{Proofs to lemma in Section~\ref{sec:LPprelim}}
\begin{repeatlemma}
[Lemma~\ref{lem:autoUB1}.]
Let the vector $V$ satisfy the following set of inequalities:
\begin{eqnarray}
\label{eq:GenIBE3}
\left[I-\lambda^LP_{\pi}^L\right]V \geq \left[I+ \lambda P_{\pi} + \cdots + \lambda^{L-1}P_{\pi}^{L-1}\right]R_{\pi} , \quad \forall \; \pi.
\end{eqnarray}
Then, we have $V \geq V^*$.
\end{repeatlemma}
\proof{Proof of Lemma~\ref{lem:autoUB1}.} 
For every stationary policy $\pi$, we have:
\begin{eqnarray}
\label{eq:iBellAllPol}
\left[I-\lambda^L P^L_\pi\right] V &\geq& \left[I + \lambda {P_\pi} + \cdots + \lambda^{L-1} P^{L-1}_\pi\right] R_\pi.
\end{eqnarray}
Since ${P_\pi}$ is a stochastic matrix (i.e., it is non-negative and its row sum equals $1$),
and $\lambda \in [0,1)$, the matrix $\left[I-\lambda^L {P_\pi}^L\right]^{-1}$ admits the following analytic series expansion:
\begin{equation}
\left[I-\lambda^L {P_\pi}^L\right]^{-1}=I+\lambda^L {P_\pi}^L +\lambda^{2L}P^{2L}_\pi+\hdots. \nonumber
\label{eq:}
\end{equation}
So, all the entries of $\left[I-\lambda^L {P_\pi}^L\right]^{-1}$ are non-negative and hence (\ref{eq:iBellAllPol}) implies the following (although the converse is not true!):
%
\begin{eqnarray}
\label{eq:upperbnd}
V &\geq& \left[I - \lambda^LP^L_\pi\right]^{-1}\left [I+ \lambda {P_\pi} + \cdots + \lambda^{L-1}P^{L-1}_\pi\right] R_\pi= \sum_{i=0}^\infty\lambda^iP^i_\pi R_\pi,\quad\forall\pi.
\end{eqnarray}
So, $V$ dominates the expected payoff associated with every policy $\pi$, including the optimal policy $\pi^*$. Hence $V\geq V^*$.
\Halmos 
\endproof


\section{Proof to lemma in Section~\ref{sec:resLP}}
\begin{repeatlemma}[Lemma~\ref{lem:surrLP}.]
Consider a surrogate LP for the $RLP$ through a set of dual variables, $\mu$ given by:
\begin{eqnarray}
\label{eq:SLP1}
SLP(\mu) &:=& \min  \bar c^T v,  \quad \mbox{subject to}\quad \\
\nonumber
\sum_{x \in {\cal S}_i} \mu_u^i(x) v(i) &\geq&  \sum_{x \in {\cal S}_i} 
\mu_u^i(x) \left[R_u(x) + \lambda \sum_{l=0}^{m} p_l v({\bar f(x,u,Y_l)})\right], \; \forall u,\; i=1,\ldots,M.
\end{eqnarray}
Then, $\exists\bar \mu \geq 0$ such that,
$SLP(\bar \mu) = RLP,$ and, for every partition index $i=1,\ldots,M$, $\exists u$ such that $\sum_{x \in {\cal S}_i}\bar \mu_u^i(x) > 0$. 
Moreover, the optimal solution $v^*$ to $RLP$ is independent of the cost vector $\bar c$ and any other feasible solution $v$ to $RLP$ dominates $v^*$.
 \end{repeatlemma}
\proof{Proof of Lemma~\ref{lem:surrLP}.} 
Consider the Langrangian dual problem to $RLP$,
$$LD(\mu) :=\min_v \left\{\bar c^T v - \sum_{i,u}\sum_{x \in {\cal S}_i}  \mu_u^i(x)\left[v(i) - R_u(x) - \lambda \sum_{l=0}^{m} p_l v({\bar f(x,u,Y_l)})\right]\right\}. $$
Let $\phi(v, \mu) = \bar c^T v - \sum_{i,u}\sum_{x \in {\cal S}_i}  \mu_u^i(x)\left[v(i) - R_u(x) - \lambda \sum_{l=0}^{m} p_l v({\bar f(x,u,Y_l)})\right]$. Let ${\mathcal F}$ be the feasible set for $RLP$ and let ${\mathcal F}(\mu)$ be the feasible set of $SLP(\mu)$. Then, we have,
\begin{eqnarray*}
LD(\mu) &:=& \min_{v} \phi(v, \mu) \\
&\leq& \min_{v \in SLP(\mu)} \phi (v, \mu) \\
&\leq& \min_{v \in SLP(\mu)}\bar c^T v = SLP(\mu).
\end{eqnarray*}
Since ${\cal F} \subset {\cal F}(\mu)$ for every $\mu$, it readily follows that $SLP(\mu) \leq RLP$. Also,
$RLP$ is feasible. For eg., consider the feasible solution $\tilde v$ given by,  
\[\tilde v(i)  = \frac{\max_{x,u}R_u(x)}{1- \lambda}, \forall i\in\{1,\ldots,M\}.\] Moreover, any feasible $v$ satisfies,
\[v(i) \geq \frac{\min_{x,u}R_u(x)}{1-\lambda}, \forall i\in\{1,\ldots,M\}.\] 
So, $RLP$ is also bounded from below and hence it satisfies the requirements of strong duality for LPs. 
Hence, there exists a $\bar \mu$ which is optimal for the dual of $RLP$ and also satisfies $LD(\bar \mu) = RLP$. Therefore, the same $\bar \mu$ must also be such that $SLP(\bar \mu) = RLP$. 
Now for every partition index $i=1,\ldots,M$, there exists at least one $u$ for which $\sum_{x \in {\cal S}_i} \bar \mu_u^i(x) > 0$. If for some $i$, $\bar \mu_u^i(x) = 0$ for every $x \in {\cal S}_i$ and for every $u$, then $SLP(\bar \mu)$ will not have any constraints lower bounding $v(i)$. It will then admit solutions for $v(i)$ that are arbitrarily negative and correspondingly, one can find a direction in which the cost of $SLP(\bar \mu)$ decreases without bound. However, this is a contradiction, since $RLP$ is lower bounded. So, we can rewrite $SLP(\bar\mu)$ in the following manner:
\begin{eqnarray}
\label{eq:SLPB}
SLP(\bar \mu) &:=& \min \bar c^T v, \quad\mbox{subject to}\quad\\
\nonumber
v(i)  &\geq& r_u(i) + \lambda \frac{1}{\sum_{x \in {\cal S}_i} \bar\mu_u^i(x)}\sum_{x \in {\cal S}_i} \bar \mu_u^i(x)\sum_{l=0}^{m} p_l v({\bar f(x,u,Y_l)}), \; \; \forall u\in\mathcal U_i,i=1,\ldots,M,
\end{eqnarray}
where, $u\in\mathcal U_i$ if $\sum_{x \in {\cal S}_i} \bar \mu_u^i(x) > 0$.
Clearly, $SLP(\bar \mu)$ is the exact LP corresponding to 
a MDP of reduced dimension with one-step reward function,
\begin{equation*}
    \label{eq:randsel1}
    r_u(i) = \frac{\sum_{x \in {\cal S}_i} \bar \mu_u^i(x)R_u(x)}{\sum_{x \in {\cal S}_i} \bar\mu_u^i(x)}, \quad \forall u\in\mathcal U_i,
   \end{equation*}
and transition probability matrix $\tilde P_u$ given by,  
  \begin{equation*}
    \label{eq:randsel1}
    \tilde P_u(i,j) :=\left\{\begin{array}{ll}
\frac{1}{\sum_{x \in {\cal S}_i} \bar\mu_u^i(x)}\sum_{x \in {\cal S}_i}\bar \mu_u^i(x) \sum_{y\in\mathcal S_j}P_u(x,y), &\mbox{ if } u\in\mathcal U_i,\\
0, &\mbox{ otherwise. }\end{array}\right.
  \end{equation*}
So, by Lemma~\ref{lem:LPopt1}, the optimal solution $v^*$ is also the optimal value function associated with the same underlying MDP. Also 
any feasible $v$ to $RLP$ is also a feasible solution to $SLP(\bar\mu)$ since the constraints for $SLP(\bar\mu)$ are obtained by a convex combination of the constraints of $RLP$. So, it follows from Lemma~\ref{lem:autoUB1} that $v \geq v^*$. 

Finally, let $RLP(\bar c)$ and $RLP(\bar d)$ denote the restricted LPs corresponding to two different cost vectors $\bar c$ and $\bar d$ respectively. Let the corresponding optimal solutions be $v_c^*$ and $v_d^*$. Since $v_d^*$ is a feasible solution for $RLP(\bar c)$, we have $v_d^*\geq v_c^*$. By the same token, $v_c^*\geq v_d^*$. Hence, $v_c^* = v_d^*$. 
\Halmos 
\endproof


\section{Proofs to claims in Section~\ref{sec:PatrolStruc}}
\begin{repeatclaim}
[Claim~\ref{cl:monoSt}.]
If ${x}_1 \geq{x}_2$, then for the same sequence of inputs $\mathbf u_t$ and disturbances $\mathbf y_t$, the system state evolves in such a way that ${x}(t;{x}_1,\mathbf u_t,\mathbf y_t) \geq {x}(t;{x}_2,\mathbf u_t,\mathbf y_t) $ for every $t > 0$.
\end{repeatclaim}
\proof{Proof of Claim~\ref{cl:monoSt}.} We use induction. Clearly at $t = 0$, ${x}_1 \geq{x}_2$. By the semi-group property of state transitions, it is sufficient to show that the result holds for $t=1$. We define the state, ${x}$, of the patrol system to be of two types.
If the following holds:
\begin{equation}
\label{eq:type1def}
\ell_x \in \Omega, \; d_x = 0, \; \mathcal A_{\ell_x,x}=1, \;\mbox{and }
\mathcal A_{j,x}=1, \mbox{ for some } j\in\Omega,j\neq \ell_x,
\end{equation}
 i.e., the UAV is at a station with an alert, the dwell time is zero and also there is an alert at some other station, then the state $x$ is of type 1. Else it is of type 2.
Note that if ${x}_1 \geq{x}_2$, then the states ${x}_1$ and ${x}_2$ are necessarily of the same type. The key property we will be using in proving Claim~\ref{cl:monoSt} is the following: service delay at a station either remains at zero (if no new alert has occurred there) or it goes up by $1$ (if there is an unserviced alert there) or it is reset to zero (if a UAV decides to loiter there).

If ${x}_1$ and ${x}_2$ are of type 1 and the UAV chooses to loiter, i.e., $u(0)=0$,  we clearly see that neither the location nor the dwell will differ at $t=1$. 
Furthermore, the delays at $t=1$ associated with the stations corresponding to initial state ${x}_1$ will be no less than the delays associated with stations corresponding to initial state ${x}_2$ since ${x}_1 \geq{x}_2$. If  $z_1={x}(1;{x}_1,0,y(0))$ and $z_2={x}(1;{x}_2,0,y(0))$,
we see that 
$\ell_{z_1}=\ell_{z_2},\;d_{z_1}=d_{z_2},\;\omega_{z_1}=\omega_{z_2},$ and $\tau_{j,z_1}\geq\tau_{j,z_2},\;\forall j\in\Omega$ for every disturbance $y(0)$ and so $z_1\geq z_2$.
The same relationship holds for other possible control choices, $u(0)\neq 0$, as well. By a similar argument, one can show that ${x}(1;{x}_1, u(0), y(0)) \geq {x}(1,{x}_2, u(0), y(0))$ holds, regardless of the control choice, even if the states ${x}_1,\;{x}_2$ are of type 2.
We use the semi-group property as follows: suppose the claim holds for all $t$ lying between $0$ and $l$ for some $l>0$. Then, we will treat the state at $t=l$ as the initial condition for determining the evolution of the state at $t= l +1$. The clock is reset as:
$\tilde t = t- l, \; \; t \geq l$. By the preceding arguments, Claim~\ref{cl:monoSt} holds for $\tilde t =1$ which is equivalent to saying that it holds for $t = l+1$.
\Halmos 
\endproof
\begin{repeatclaim}
[Claim~\ref{cl:monoVal}.]
If ${x}_1 \geq{x}_2$, then $V^*({x}_1) \leq V^*({x}_2)$. Furthermore, if $\mathcal S_i\geq\mathcal S_j$, then $\min_{{x} \in {\cal S}_i} V^*({x}) \leq \min_{{z} \in {\cal S}_j} V^*({z})$.
\end{repeatclaim}
\proof{Proof of Claim~\ref{cl:monoVal}.} Let $\pi^*$ be the optimal policy; accordingly $\pi^*({x})$ is fixed for every $x\in\mathcal S$. Then, for every $t > 0$, we can determine ${x}(t;{x}_1, \mathbf u^*_t, \mathbf y_t)$  for some sequence of disturbances $\mathbf y_t$, where the optimal input sequence $\mathbf u_t^*=\{u^*(0),\ldots,u^*(t-1)\}$ (starting with $x_1$) can be recursively obtained as follows: 
\begin{equation}
\label{eq:optconx1}
u^*(t) = \pi^*({x}(t-1;{x}_1,\mathbf u_{t-1}^*, \mathbf y_{t-1})).
\end{equation}
with the initialization $u^*(0) = \pi^*(x_1)$. For the above $\mathbf u^*$ and $\mathbf y$, we can then determine the evolution of the states corresponding to initial state ${x}_2$. Since ${x}(t;{x}_1,\mathbf  u_t^*,\mathbf  y_t) \geq {x}(t;{x}_2,\mathbf  u_t^*,\mathbf  y_t)$ by Claim~\ref{cl:monoSt}, we notice readily that the reward
$R_{u^*}({x}(t;{x}_1,\mathbf  u_t^*,\mathbf  y_t)) \leq R_{u^*}({x} (t;{x}_2,\mathbf  u_t^*,\mathbf  y_t))$ for every $t \geq 0$ (since the one-step reward is based only on the maximum delay, dwell time and control input, the inequality follows). Since the above holds for any given disturbance sequence, the expected discounted payoff associated with the state starting from ${x}_1$ i.e., $V^*(x_1)$, is no more than the expected discounted payoff associated with the state starting from ${x}_2$, which we will denote by $V_{\mathbf u^*}(x_2)$. As a result, $V^*({x}_1) \leq V_{\mathbf u^*}(x_2)\leq V^*({x}_2)$. The second part of the inequality holds since $\mathbf u_t^*$ as defined in (\ref{eq:optconx1}) is a sub-optimal control policy for the state evolution starting from $x_2$ and hence the expected discounted payoff associated with that policy is necessarily dominated by the optimal value function starting from $x_2$.
To complete the proof, consider two different partitions $\mathcal S_i$ and $\mathcal S_j$ such that $\mathcal S_i \geq \mathcal S_j$. Let $\bar{z} =\argmin_{{z} \in {\cal S}_j} V^*({z})$ and this can always be found since we are dealing with a subset, ${\cal S}_j$ of a finite state space $\mathcal S$. Since $\mathcal S_i \geq \mathcal S_j,\;\exists {\bar x} \in {\cal S}_i$ such that ${\bar x} \geq \bar{z}$. We have shown that for this case, $V^*({\bar x}) \leq V^*(\bar{z}) = \min_{{z} \in {\cal S}_j} V^*({z})\Rightarrow\min_{{x}\in {\cal S}_i} V^*({x}) \leq \min_{{z} \in {\cal S}_j} V^*({z})$.
\Halmos 
\endproof
\ACKNOWLEDGMENT{}
This work was also partly supported by the AFRL Summer Faculty Program and AFOSR award no. FA9550-10-1-0392.




%
\bibliographystyle{nonumber}

\end{document}